\documentclass{aa}
\usepackage{psfig,times}

\usepackage{graphicx}

\newcommand\ra[4]{~~#1$^{\rm h}$#2$^{\rm m}$#3$^{\rm s}$.#4 }
\newcommand\dec[4] {$#1$$^{\circ}$#2$^{\rm '}$#3$^{\rm ''}$.#4}

\newcommand{\sbb}{mag/$\sq\arcsec$}

\def\ha{H$\alpha$}

\def\h2{H{\small II}}
\newcounter{qub}
\def\sbu{mag/arcsec$^2$}

\def\p23{{\it P$^{\tiny \cal J}_{23}$}}
\def\e25{{\it E$^{\cal J}_{23}$}}

\def\msun{$M_{\odot}$}
\def\zsun{$Z_{\odot}$}

\def\ha{H$\alpha$}

\def\rr{$R^*$}

\def\kmsec{${\rm km\,\,s^{-1}}$}

\def\flat{type\,V}
\def\med{{\sf med }}
\def\ser{S\'ersic }
\def\zsun{$Z_{\odot}$}

\begin{document}

\title{New insights to the photometric structure of Blue Compact Dwarf 
Galaxies from deep Near-Infrared studies:}
\subtitle{II. The sample of northern BCDs\thanks{Table 2 is also available in electronic form at the
CDS via anonymous ftp to cdsarc.u-strasbg.fr (130.79.128.5) or via
http://cdsweb.u-strasbg.fr/cgi-bin/qcat?J/A+A/}}

\author{K.G. Noeske\inst{1}\inst{2}
\and P. Papaderos\inst{1}
\and L.M. Cair\'os\inst{1}
\and K.J. Fricke\inst{1}}
\offprints{kai@ucolick.org}
\institute{      Universit\"ats--Sternwarte, Geismarlandstra\ss e 11,
                 D--37083 G\"ottingen, Germany
\and Lick Observatory, Univ. of California, 1156 High Street, Santa Cruz, CA 95064, USA}
\date{Received \hskip 2cm; Accepted}

\abstract{This paper is part of a series of publications which present
a systematic study of Blue Compact Dwarf (BCD) Galaxies in the Near
Infrared (NIR). Compared to the visible light, NIR data allow a
better separation of the starburst emission from the light
distribution of the old stellar low-surface brightness (LSB) host
galaxy.
We analyze deep NIR broad band images of a sample of 11 BCDs,
observed with the Calar Alto
\thanks{German--Spanish Astronomical Center, Calar Alto, operated
by the Max--Planck--Institute for Astronomy, Heidelberg, jointly with
the Spanish National Commission for Astronomy.} 
3.6m telescope.
This work enlarges the samples presented in preceding papers of this study
(Noeske et al. 2003, Cair\'os et al. 2003) by BCDs of the most
common morphological type, displaying a regular elliptical LSB host
galaxy.
The data presented here allow the detection and quantitative study
of the extended stellar LSB host galaxy in
all sample BCDs. The NIR surface brightness profiles (SBPs) of the LSB
host galaxies agree at large galactocentric radii with those from
optical studies, showing also an exponential intensity decrease and
compatible scale lengths.
Similar to Noeske et al. (2003), we find centrally flattening
exponential (\flat ) SBPs of the host galaxy for several BCDs.  Such
SBPs remain mostly undetected in optical bands, due to the comparatively
stronger starburst emission at these wavelengths. We apply a modified
exponential distribution to decompose and quantitatively analyze SBPs
of LSB hosts with a \flat\ intensity distribution.
We present the results of the surface photometry and the decomposition
of SBPs, and discuss
individual objects with respect to morphological details of their
star-forming regions.}

\authorrunning{Noeske et al.}
\titlerunning{NIR observations of blue compact dwarf galaxies II}

\maketitle

\section{Introduction\label{intro_nir2}}
\begin{table*}
\caption{Sample galaxies; see Sect. \ref{data_nir2}}
\label{tab_sample_nir2}
\tabcolsep2mm
\begin{tabular}{llcccccccl}\hline\hline
Object       &   RA(J2000)      & t$_{J}$ & t$_{H}$ &t$_{K'}$& M$_B$  & A$_B$&    D  &FWHM                  & other       \\
             &   DEC(J2000)     &   	   & 	     &         &(ref.)&      &(ref.) &(final)$^{\spadesuit}$& names       \\
             &                  &  [s]     & [s]     & [s]     & [mag]& [mag]&[Mpc]  &[\arcsec]	     &             \\
             &   (2)            & (3)      & (4)     & (5)     & (6)  & (7)  &   (8) &    (9)  	     & (10)         \\ \hline
Mkn 314      &\ra{23}{02}{59}{3}& 1440     & 1740    & 1920    &-18.5 & 0.38 &   28.9&   1.5   	     & NGC 7468;     \\
(iE)         &\dec{+16}{36}{18}{9}&        &         &         & (f)  &       & (a)   &       	     & UGC 12329 \\ \hline\hline
Mkn 209      &\ra{12}{26}{16}{0}& 1620     & 1440    & 1140    &-14.2 & 0.07 &   5.8 &   1.5         & UGCA\,281; I\,Zw\,36;    \\
(iE)         &\dec{+48}{29}{36}{6}&        &         &         & (g)  &      & (b)   &       	     & Haro\,29		\\ \hline\hline
Mkn 996      &\ra{01}{27}{35}{5}& 1260     & 1440    & 1200    &-16.6 & 0.19 &  20.4 &   1.4         &     \\
(nE)         &\dec{-06}{19}{36}{1}&        &         &         & (h)     &      & (c)   &       	     &		\\ \hline\hline
Mkn 370      &\ra{02}{40}{29}{0}& 1140     & 1860    & 1640    &-17.1 & 0.40 &  11.2 &   1.2         & NGC\,1036; IC\,1828;    \\
(nE)         &\dec{+19}{17}{49}{6}&        &         &         & (f)  &      & (a)   &       	     & UGC\,2160 \\  \hline\hline
I Zw 115     &\ra{15}{32}{57}{0}&  672     & 672     & 480     &-16.4 & 0.06 &  15.0 &   1.3         & UGC\,9893;    \\
(iI)         &\dec{+46}{27}{06}{5}&        &         &         & (g)  &      & (a)   &       	     & VV\,720 \\  \hline\hline
Mkn 5        &\ra{06}{42}{15}{5}&  1650    & 2100    & 2340    &-15.7 & 0.36 &  15.3 &   1.5         & UGCA 130    \\
(iE/iI,C)    &\dec{+75}{37}{32}{6}&        &         &         & (f)  &      & (a)   &       	     & 		\\  \hline\hline
Mkn 600      &\ra{02}{51}{04}{6}&  1560    & 1320     & 1080   &-15.5 & 0.28 &  12.6 &   1.1         &     \\
(iE)	     &\dec{+04}{27}{13}{9}&        &         &         & (f)  &      & (a)   &       	     & 		\\  \hline\hline
NGC 6789     &\ra{19}{16}{41}{9}&  1200    & 1680     & 1920   &-14.3 & 0.30 &  3.6  &   1.4         & UGC 11425    \\
(iE)	     &\dec{+63}{58}{18}{0}&        &         &         & (i)  &      & (d)   &       	     & 		\\  \hline\hline
Mkn 324      &\ra{23}{26}{32}{8}&  1680    & 600      & 1140   &-16.5 & 0.22 &  21.8 &   1.4         & UGCA 439    \\
(iE)	     &\dec{+18}{15}{59}{8}&        &         &         & (f)  &      & (a)   &       	     & 		\\  \hline\hline
Mkn 450      &\ra{13}{14}{48}{3}&  1200    & 1440     & 1920   &-16.7 & 0.06 &  17.9 &   1.3         & UGC\,08323; VV\,616;  \\
(iE)	     &\dec{+34}{52}{51}{3}&        &         &         & (j)  &      & (a)   &       	     & HS\,1312+3508 \\  \hline\hline
NGC 5058     &\ra{13}{16}{52}{3}&  720     & 1680     & 1440   &-15.9 & 0.13 &  11.6 &   1.1         & UGC\,08345; Mkn\,786;    \\
(iI?)	     &\dec{+12}{32}{53}{9}&        &         &         & (i)  &      & (c,e)&		     & KPG\,370 \\  \hline\hline
\end{tabular}\\
$\spadesuit$: Resolution of the best image set available for the
respective galaxy, after reduction and combination (a):
Tully (\cite{tully88}), (b): Schulte--Ladbeck et al. \cite{schulte01}, (c): inferred from
the heliocentric velocity $v_{\rm hel}$ listed in the NED, corrected
for solar motion with respect to the center of the Virgo Cluster and
adopting H$_0$=75\,km\,s$^{-1}$\,Mpc\,$^{-1}$ (cf. \cite{n03a}), (d):
Drozdovsky et al. (\cite{drozdovsky01}), (e): see 
Sect. \ref{n5058}, (f): m$_B$ from Table 2 in Cair\'os et al. 2001a,
(g): m$_B$ from Papaderos et al. 1996a, (h): m$_B$ from Thuan et
al. \cite{thuan96}, (i): m$_B$ from the RC3 (de Vaucouleurs et
al. \cite{devaucouleurs91}), (j): m$_B$ from Vennik et
al. \cite{vennik00}.\\ \hrule
\end{table*}

This paper is part of a series of publications which present a deep
Near Infrared (NIR) imaging and surface photometry study of a large
sample of Blue Compact Dwarf (BCD) Galaxies (Cair\'os et al. 2003,
hereafter \cite{c03a}, Noeske et al. 2003, hereafter \cite{n03a}).
As described in these papers, this project aims at an improved
understanding of the structure and photometric properties of the old
stellar low-surface brightness (LSB) host galaxy, and of the young
stellar populations in BCDs.  The LSB component contains practically
all stellar mass of typical BCDs, and is therefore a likely
dynamically important constituent of such galaxies (see references in
\cite{n03a}). Its radial light distribution, i.e. its projected
luminosity density distribution, provides close constraints to the
stellar mass distribution in a BCD, information which is crucial for
modelling the global gravitational potential and dynamics of BCDs, as well
as the effects of starburst events in such objects, such as galactic
winds (e.g. De Young \& Heckman \cite{deyoung94}, Mac Low \& Ferrara
\cite{maclow99}, Silich \& Tenorio-Tagle \cite{silich01}). The
structural parameters (e.g., exponential scale length and central
surface brightness) of the stellar LSB host also form a prime
diagnostic tool for assessing the proposed evolutionary relations
between different types of dwarf galaxies (see e.g. Lin \& Faber
\cite{lin83}, Thuan \cite{thuan85}, Dekel \& Silk \cite{dekel86},
Davies \& Phillipps \cite{davies88}, Papaderos et al. 1996b, hereafter
\cite{papaderos96b}, Marlowe et al. \cite{marlowe99}, Dekel \& Woo
\cite{dekel03}, see also the references in \cite{n03a}). Of equal
interest are the relations between the properties of the stellar host
galaxy and the occurrence and properties of starburst activity in BCDs
(Loose \& Fricke \cite{loose82}, \cite{papaderos96b}). The
investigation of these latter issues requires a separation of the light
distributions of the young and old stellar populations by means of a
decomposition of radial surface brightness profiles (SBPs). Such
analyses are however hampered by the extended dominant starburst
emission at visible wavelengths, and can be performed with much
better precision in the NIR (see \cite{n03a}, \cite{c03a}).

In the current paper, we present $J$, $H$ and $K'$ image data and
surface photometry of a sample of 11 BCDs, as well as a decomposition
of the derived SBPs into the radial intensity distributions of the old
LSB host galaxy and the starburst component.  Morphological
information is provided for each object, in particular for the
star-forming (SF) regions. Remarkable features are briefly
discussed. All data reduction and analysis procedures that were
described in \cite{n03a} have been unalteredly applied to the present
data set.

Most of the objects in the present sample show an iE/nE morphology,
according to the classification scheme by Loose \& Thuan (1986,
hereafter \cite{loose86}). Such BCDs display SF regions
either irregularly distributed (iE) or centrally confined (nE) within
a smooth elliptical or circular stellar LSB host galaxy. The iE/nE
BCDs comprise the majority ($\sim 80-90\%$) of the BCDs in the local
Universe (\cite{loose86}), but were underrepresented in the sample
previously studied by \cite{n03a} which comprises mainly irregular
morphological types and interacting objects. Earlier studies have
indicated that the various morphological types of BCDs may
systematically differ from each other in several physical properties
(e.g. Salzer et al. \cite{salzer89b}, Telles et al. \cite{telles97b},
Noeske et al. \cite{noeske00}).  For later comparisons of their NIR
properties, a balanced representation of the different morphological types
is therefore desirable. Analyses of the cumulative sample will be
presented in forthcoming papers of the present series.

This paper is structured as follows: in Sect. \ref{data_nir2}, we list
the subsample under study, and summarize the observations. In
Sect. \ref{deco_nir2}, some details concerning the derivation,
analysis and decomposition of SBPs, described in depth in
\cite{n03a}, are repeated to facilitate the understanding of this
paper. This section also lists the results of the surface photometry, and
of the decomposition of the surface brightness profiles, as well as
colors of the old stellar host galaxies.  Individual objects are
presented and briefly discussed in Sect. \ref{sample_galaxies_nir2},
along with images, surface brightness profiles and color profiles. 
A brief summary is given in Sect. \ref{summary}.

\section{Observations and data reduction}
\label{data_nir2}

The galaxies of this sample are listed in Table \ref{tab_sample_nir2},
along with their adopted distances, absolute $B$ magnitudes, Galactic
extinction, and references to literature sources from which the latter
values were taken. Selection criteria for distance determinations
from the literature and details on the Galactic extinction values we
adopt are given in Sects. 2.1 and 2.4.2 of \cite{n03a},
respectively.

The NIR images were observed with the 3.6m telescope of the
German--Spanish Astronomical Center, Calar Alto, Spain. 
Atmospheric conditions during the different observing runs were
generally satisfactory (April 3rd, 1999: FWHM 1\farcs 5, transparency
average; December 26th, 1999: FWHM 1\farcs 3 -- 3\farcs 5,
transparency fair; May 10th--15th, 2000: FWHM 0\farcs 8 -- 1\farcs 4,
transparency good to average; October 6th--10th, 2000: FWHM 1\farcs 2
-- 2\farcs 0, transparency good to fair).
The telescope was equipped with the OMEGA PRIME camera, mounted at the
prime focus. The 1024$\times$1024 pixel Rockwell HAWAII detector of
this instrument provided a pixel scale of 0\farcs 396 and a field of view
of 6\farcm 76\ .
Images were taken through the broad band filters $J$ and $H$, as well
as the $K$\arcmin\ filter, which was preferred to a normal $K$ filter
to attenuate the contribution of thermal background. Information on
the observing technique and control of the time-dependent NIR
background emission can be found in \cite{n03a}.
The total on-object exposure times for each galaxy, after rejection of
subexposures that were affected by unstable readout electronics or
strong background gradients, are listed in columns 3 -- 5 of Table
\ref{tab_sample_nir2}.
All data reduction procedures that were applied to the NIR images are
detailed in \cite{n03a}. The angular resolution of the resulting
images is listed in column 9 of Table \ref{tab_sample_nir2}. 
%
\subsection{Flux calibration and extinction correction}
\label{cal_nir2}

The transparency variations within each night, $\ga 0.15$ mag in the
$K'$ band, did not allow a precise flux calibration through
observations of standard stars.  All data were therefore calibrated
using fluxes of bright (typically 10 ... 15 $J$ mag) field stars in
the vicinity of the sample galaxies, given in the {\em Two Micron All
Sky Survey} (2MASS)
catalogue\footnote{http://www.ipac.caltech.edu/2mass/} (see Cutri et
al. \cite{cutri00}, Jarrett et al. \cite{jarrett00}). { Calibration
errors for each galaxy are shown in column 14 of Table
\ref{tab_phot_nir2}. These include uncertainties of 2MASS photometry
for the field stars used for calibration, and measurement errors of
these stars in our images.} To calibrate the $K$\arcmin\ images, the
2MASS $K_s$ fluxes were first transformed to $K$\arcmin\ magnitudes
using Eqs. (3) from \cite{n03a}.  The results listed within this paper
therefore refer to the $J$,$H$,$K$\arcmin\ photometric system defined
by the Calar Alto 3.6m telescope and the OMEGA PRIME camera, tied to
2MASS zero points.  Since color terms of the transformation between
either photometric system are not available, intrinsic uncertainties
of several 0.01 mag may be present. Notes on these uncertainties and
on comparisons to data calibrated in other photometric systems are
given in Sect. 2.4.1 of \cite{n03a}, and in Sect. \ref{lsbcolors_nir2}
of this paper.

Magnitudes and colors given in this paper are corrected for Galactic
extinction, adopting values (see column 7 of Table
\ref{tab_sample_nir2}) derived from the $B$ band extinction maps by
Schlegel et al. (\cite{schlegel98}) (cf. Table \ref{tab_sample_nir2})
and the standard (R$_V=3.1$) extinction law (Cardelli et
al. \cite{cardelli89}) implemented into the NED.  No attempt was made
to correct for internal extinction, since this is known to vary
spatially even in the most metal-deficient BCDs (cf., e.g., Guseva et
al. \cite{guseva01}, Cannon et al. \cite{cannon02}) and can be
reliably constrained in the SF regions only.

\section{Surface photometry and profile decomposition}   
\label{deco_nir2}
\begin{table*} 
\tabcolsep1.5mm
\caption{Structural properties of the dwarfs$^a$; see Sect. \ref{deco_nir2} for explanations. See also discussions of individual objects.}
\label{tab_phot_nir2}
\begin{tabular}{llcccccccccccc}
\hline
\hline
Name & Band & $(\mu_{E,0})^c$ & $\alpha $ & $m_{\rm LSB}^{\rm fit}$ &
$P_{\rm iso}$  & $m_{P_{\rm iso}}$ & $E_{\rm iso}$ & $m_{\rm E_{\rm iso}}$  & $m_{\rm SBP}$ &
$(m_{\rm tot})^c$ & $r_{\rm eff}$,$r_{80}$ & $\eta _{\rm\,SBP}$ & $\sigma _{cal}$ \\
($b$,$q$)$^b$&     & \sbb\ &  kpc   & mag    & kpc       &  mag       &  kpc
          &  mag             &    mag & mag & kpc & & mag\\
              &   &       &       & & & & & & & & & & \\ 
   (1) &   (2)         &   (3)     &  (4)      &   (5)            &
 (6)    &  (7)             &  (8)     &  (9)  & (10) & (11) & (12) & (13) & (14) \\
\hline
Mkn\,314     	&$J$ & 19.10$\pm$0.13 & 1.00$\pm$0.04  & 12.84 & 1.85 & 12.84 & 3.58 & 12.99 & 12.12 & 12.05$\pm$0.01 & 0.88,1.86 & 1.46 & 0.03\\
$\star^f$         &$H$ & 18.61$\pm$0.15 & 1.09$\pm$0.05  & 12.17 & 1.70 & 12.26 & 3.39 & 12.39 & 11.51 & 11.38$\pm$0.01 & 0.90,1.96 & 1.50 & 0.03\\
               	&$K'$& 18.73$\pm$0.17& 1.22$\pm$0.07& 12.04 & 2.12 & 11.77 & 3.67 & 12.28 & 11.21 & 11.17$\pm$0.02 & 0.88,1.88 & 1.45 & 0.03\\
\hline
Mkn\,209     	&$J$ & 20.21$\pm$0.06 & 0.23$\pm$0.01  & 13.63 & 0.16 & 16.37 & 0.60 & 13.98 & 13.69 & 13.48$\pm$0.01 & 0.29,0.47 & 1.69 & 0.03\\
$\star^f$       &$H^{d,i}$ & --- & --- & --- & --- & --- & --- & --- & 13.38 & 13.15$\pm$0.01 & 0.29,0.45 & 1.31 & 0.06\\
               	&$K'^h$& --- & --- & --- & --- & --- & --- & --- & 13.10 & 13.01$\pm$0.06 & 0.28,0.44 & 1.37 & 0.07\\
\hline
Mkn\,996     	&$J$ & 18.20$\pm$0.03 & 0.40$\pm$0.01  & 13.17 & 0.67 & 14.50 & 1.77 & 13.24 & 12.91 & 12.87$\pm$0.01 & 0.51,1.01 & 1.53 & 0.03\\
$\star$         &$H$ & 17.80$\pm$0.04 & 0.42$\pm$0.01  & 12.65 & 0.69 & 13.74 & 1.64 & 12.76 & 12.33 & 12.26$\pm$0.01 & 0.51,1.03 & 1.67 & 0.04\\
               	&$K'^e$&17.48$\pm$0.05& 0.43$\pm$0.01  & 12.32 & 0.68 & 13.20 & 1.77 & 12.41 & 12.04 & 11.94$\pm$0.02 & 0.42,0.84 & 1.90 & 0.05\\
\hline
Mkn\,370     	&$J$ & 18.85$\pm$0.05 & 0.64$\pm$0.01  & 11.52 & 0.80 & 12.80 & 2.43 & 11.64 & 11.29 & 11.27$\pm$0.01 & 0.72,1.49 & 1.68 & 0.03 \\
$\star^g$         &$H$ & 18.36$\pm$0.06 & 0.63$\pm$0.01  & 11.04 & 0.84 & 12.14 & 2.12 & 11.22 & 10.74 & 10.72$\pm$0.04 & 0.71,1.46 & 1.72 & 0.07\\
               	&$K'$& 18.00$\pm$0.22 & 0.59$\pm$0.04  & 10.83 & 0.78 & 12.03 & 2.17 & 10.97 & 10.57 & 10.46$\pm$0.03 & 0.67,1.39 & 1.55 & 0.05\\
\hline
I\,Zw\,115     	&$J$& 19.52$\pm$0.13 & 0.53$\pm$0.02  & 13.68 & 0.85 & 15.05 & 1.68 & 14.01 & 13.46 & 13.26$\pm$0.02 & 0.88,1.46 & 1.15 & 0.04\\
$(2.3,0.8)$     &$H$& 19.23$\pm$0.35 & 0.56$\pm$0.06  & 13.25 & 0.80 & 14.49 & 1.35 & 13.91 & 13.00 & 12.89$\pm$0.07 & 0.88,1.46 & 1.15 & 0.06\\
               	&$K'^h$& ---           & ---            & ---   & ---  & ---   & ---  & ---   & 12.87 & 12.64$\pm$0.07 & 0.80,1.32 & 1.13 & 0.08\\
\hline
Mkn\,5     	&$J$ & 19.03$\pm$0.04 & 0.37$\pm$0.01 & 13.73 & 0.54 & 15.90 & 1.35 & 13.90 & 13.73 & 13.58$\pm$0.01 & 0.56,0.92 & 1.08 & 0.03\\
$(1.4,0.65)$    &$H$ & 18.78$\pm$0.07 & 0.40$\pm$0.01 & 13.30 & 0.58 & 14.94 & 1.18 & 13.56 & 13.21 & 13.16$\pm$0.02 & 0.57,0.92 & 1.04 & 0.04\\
               	&$K'$& 18.41$\pm$0.06 & 0.40$\pm$0.01 & 12.92 & 0.56 & 14.86 & 1.33 & 13.14 & 12.93 & 12.78$\pm$0.02 & 0.57,0.91 & 1.22 & 0.05\\
\hline
Mkn\,600     	&$J$ & 19.35$\pm$0.08 & 0.30$\pm$0.01 & 13.93 & 0.41 & 15.63 & 0.99 & 14.11 & 13.86 & 13.78$\pm$0.04 & 0.35,0.64 & 1.34 & 0.05\\
$\star^f$    	&$H$ & 19.09$\pm$0.08 & 0.32$\pm$0.01 & 13.48 & 0.40 & 15.07 & 0.87 & 13.79 & 13.41 & 13.23$\pm$0.07 & 0.36,0.66 & 1.35 & 0.07\\
               	&$K'^d$& 19.01$\pm$0.25 & 0.32$\pm$0.04 & 13.42 & 0.43 & 14.67 & 0.88 & 13.72 & 13.25 & 13.13$\pm$0.07 & 0.35,0.65 & 1.52 & 0.06\\
\hline
NGC\,6789     	&$J$ & 18.79$\pm$0.13 & 0.20$\pm$0.01 & 11.93 & 0.31 & 13.96 & 0.75 & 12.11 & 11.82 & 11.78$\pm$0.03 & 0.33,0.57 & 1.07 & 0.03\\
$(3.3,0.70)$    &$H$ & 18.37$\pm$0.08 & 0.20$\pm$0.01 & 11.51 & 0.25 & 13.56 & 0.64 & 11.78 & 11.38 & 11.32$\pm$0.03 & 0.33,0.56 & 1.13 & 0.03\\
               	&$K'$& 18.12$\pm$0.32 & 0.20$\pm$0.02 & 11.25 & 0.23 & 13.58 & 0.69 & 11.48 & 11.19 & 11.08$\pm$0.04 & 0.34,0.57 & 1.26 & 0.02\\
\hline
Mkn\,324     	&$J$ & 17.81$\pm$0.03 & 0.28$\pm$0.01 & 13.68 & 0.57 & 14.61 & 1.35 & 13.73 & 13.32 & 13.25$\pm$0.01 & 0.34,0.67 & 1.26 & 0.04\\
$\star$    	&$H$ & 17.36$\pm$0.10 & 0.29$\pm$0.01 & 13.17 & 0.55 & 14.08 & 1.24 & 13.26 & 12.81 & 12.75$\pm$0.02 & 0.34,0.68 & 1.30 & 0.04\\
               	&$K'^d$&16.93$\pm$0.12& 0.26$\pm$0.01 & 12.94 & 0.50 & 14.07 & 1.24 & 13.00 & 12.66 & 12.55$\pm$0.02 & 0.32,0.62 & 1.23 & 0.05\\
\hline
Mkn\,450     	&$J$ & 19.91$\pm$0.04 & 0.87$\pm$0.01 & 13.09 & 1.41 & 14.41 & 2.48 & 13.43 & 12.90 & 12.80$\pm$0.01 & 1.25,2.08 & 1.46 & 0.05\\
$(1.5,0.65)$    &$H^d$&19.17$\pm$0.13 & 0.80$\pm$0.03 & 12.55 & 1.18 & 14.18 & 2.08 & 12.96 & 12.37 & 12.35$\pm$0.01 & 1.22,2.01 & 1.06 & 0.07\\
               	&$K'^h$& ---           & ---            & ---   & ---  & ---   & ---  & ---   & 12.16 & 12.05$\pm$0.02 & 1.20,1.99 & 1.27 & 0.05\\
\hline
NGC\,5058     	&$J$ & 19.08$\pm$0.13 & 0.47$\pm$0.02 & 12.47 & 0.72 & 13.67 & 1.70 & 12.62 & 12.21 & 12.18$\pm$0.01 & 0.56,1.13 & 1.44 & 0.05\\
$\star^f$    	&$H$ & 18.95$\pm$0.05 & 0.52$\pm$0.01 & 12.14 & 0.80 & 12.71 & 1.45 & 12.42 & 11.65 & 11.67$\pm$0.02 & 0.55,1.10 & 1.42 & 0.07\\
               	&$K'$& 18.62$\pm$0.14 & 0.49$\pm$0.02 & 11.94 & 0.79 & 12.56 & 1.52 & 12.15 & 11.49 & 11.48$\pm$0.03 & 0.52,1.02 & 1.46 & 0.07\\
\hline
\hline
\end{tabular}
\parbox{17.2cm}{$a$: All values are corrected for Galactic extinction, 
adopting the $A_B$ from Table. \ref{tab_sample_nir2}.}
\parbox{17.2cm}{$b$: See Sect. \ref{deco_nir2} for details. Objects for which the
LSB component was modelled by a pure exponential (Eq. \ref{exponential_nir2})
are marked with an asterisk.}
\parbox{17.2cm}{ $c$: Errors do not include the calibration uncertainties given in column (14).}
\parbox{17.2cm}{$d$: Decomposition less reliable in this filter (low $S/N$, or data affected by nearby bright stars).}
\parbox{17.2cm}{$e$: $K'$ SBP less deep than $J$ and $H$; effects on columns (10), (11) \& (13).}
\parbox{17.2cm}{$f$: small systematic deviations from an exponential LSB profile, likely type V SBP { (see notes of caution in Sect. \ref{deco_nir2})}.}
\parbox{17.2cm}{$g$: small systematic deviations from an exponential LSB profile, possible type V SBP { (see Sect. \ref{mkn370}, and notes of caution in Sect. \ref{deco_nir2})}.}
\parbox{17.2cm}{ $h$: detected or likely type V SBP restricts exponential part of LSB fit to low $S/N$ region, preventing robust decomposition in the $K$ band.}
\parbox{17.2cm}{ $i$: likely type V SBP, cannot be reliably modelled; exponential LSB fit yields unphysical $H$ band decomposition (see Sect. \ref{mkn209}). }
\end{table*}


The SBPs and color profiles were derived as detailed in \cite{n03a},
using throughout the algorithm ``iv'' presented in Papaderos et
al. (\cite{papaderos02}).  { A comparison to SBPs derived by
alternative methods was performed for each profile to ensure
consistency at both low and high $S/N$ levels. These comparison SBPs
were generally obtained through methods ``i'' (ellipse fits to
isophotes) and ``iii'' (area of all image pixels above a given
intensity threshold) described in Papaderos et al. 1996a (hereafter
\cite{papaderos96a})} .

As described in \cite{n03a}, uncertainties as
displayed for each SBP are very conservative upper limits to the true
uncertainties and are likely often overestimated.

Also the decomposition of the derived SBPs into the intensity
distributions of the LSB host and the starburst component, which is
required to separately study the emission of either stellar component,
is explained in Sect. 3.2 of \cite{n03a}.

For a better understanding of the results of the present work (Table
\ref{tab_phot_nir2}), we reiterate here the functions that
were employed to fit the SBPs of the LSB host galaxies. The resulting
parameters of those fits provide the structural parameters of the LSB
host, and make it possible to calculate a model of its intensity distribution,
which is then subtracted from the total SBPs to obtain the light
distribution of the starburst component.

At large galactocentric radii $(> 2\dots 3$ scale lengths $\alpha )$,
the SPBs of the LSB hosts show for most BCDs a smooth decay, which can
typically be approximated by an exponential law (e.g. LT86, P96a,
Telles et al. \cite{telles97b}, Cair\'os et al. 2001a, hereafter
C01a).

If the intensity is expressed in terms of the surface
brightness $\mu$, this function reads as
\begin{equation}
\label{exponential_nir2}
\mu(R^{\star})=\mu_{\rm E,0}+1.0857\frac{R^{\star}}{\alpha}
\end{equation}
with $\mu_{\rm E,0}$, $\alpha$ and $R^{\star}$ denoting, respectively,
the extrapolated central surface brightness in \sbb , the exponential
scale length, and the photometric radius.

As detailed in \cite{n03a}, an exponential fit to the outer SBP of the
LSB host provided in some cases no meaningful decomposition, but
exceeded the intensity of the observed {\em total} SBP at small
radii. In these cases, where a central flattening of the outer
exponential intensity distribution of the LSB host (``\flat\ SBP'',
Binggeli \& Cameron \cite{binggeli91}, see Sect. 3.2 in \cite{n03a}) had to be postulated,
a modified exponential distribution Eq. (\ref{med_nir2}) was applied
(cf. \cite{papaderos96a}) to decompose the total SBPs:
\begin{equation}
\label{med_nir2}
I(R^{\star})=I_{\rm E,0}\exp
\left(-\frac{R^{\star}}{\alpha}\right)\{1-q\exp (-P_3(R^{\star}))\}
\end{equation}
with
\begin{equation}
P_3(R^{\star})=\left(\frac{R^{\star}}{b\alpha}\right)^3+
\left(\frac{R^{\star}}{\alpha}\cdot \frac{1-q}{q}\right).
\end{equation}
This empirical function, in the following referred to as \med,
flattens with respect to a pure exponential law inside a cutoff
radius $b\alpha$, and attains at \rr=0\arcsec\ an intensity given by
the relative depression parameter $q=\Delta I/I_{\rm E,0}< 1$, where
$I_{\rm E,0}$ is the extrapolated central intensity of the unflattened
exponential. Details on the determination of the parameters of the
\med, $\mu_{\rm E,0}$, $\alpha$, and $(b,q)$, can be found in
Sect. 3.2 of \cite{n03a}.

For several objects, the SBPs of the LSB host show small systematic
deviations from an exponential slope, yet at low significance
levels. { Confirmation or reliable quantification of the type V SBP
was not possible, so these galaxies were decomposed by means of
exponential fits. Nevertheless, \flat\ SBPs are likely for Mkn 209,
Mkn 314, Mkn 600 and NGC 5058, and possible for Mkn 370. If type V
SBPs of the LSB host are present in these galaxies, the decompositions
by means of exponential fits wibll be imprecise, as detailed in
\cite{n03a}. Most importantly, the resulting host galaxy flux will be
overestimated, and the starburst component correspondingly
underestimated.}

Alternative fits to the light distribution of the LSB host using the
S\'ersic law were abandoned in the present work. As detailed in
\cite{n03a} and \cite{c03a}, the use of the latter empirical function
to model the LSB emission of BCDs entails several problems and, as it
demands a currently unavailable data quality, frequently does not
yield a robust solution.  { No LSB host showed a SBP that gave signs
of being steeper than an exponential (S\'ersic index $\eta >1$).}

S\'ersic fits were merely used to describe the {\em total} SBPs. In
the case of BCDs, total SBPs include the emission of both the
starburst component and the LSB host galaxy, whose mass-to-light
ratios differ strongly from each other. These S\'ersic fits are
therefore of limited use to understand the physics of BCDs, and were
performed to allow comparisons with previous surface photometry
studies where such fits were performed, and to high-redshift objects
for which only integral profiles can be derived.

The photometric quantities that were derived for the sample BCDs are
summarized in Table \ref{tab_phot_nir2}. { BCDs for which a \flat\
profile in their underlying LSB component was either not detected, or
could not be reliably confirmed and modelled (see above), were}
decomposed by means of a pure exponential (Eq. \ref{exponential_nir2})
and are marked with an asterisk in column 1. For the remaining systems
we list the ($b$,$q$) parameters, as obtained by fitting
Eq. (\ref{med_nir2}) to SBPs of the LSB host galaxy. The latter SBPs were
derived from images out of which irregular starburst emission had been
largely removed (\cite{n03a}).
Columns 3 and 4 list, respectively, the \emph{extrapolated} central
surface brightness $\mu_{\rm E,0}$ and the exponential scale length
$\alpha$, obtained by fitting an exponential
(Eq. \ref{exponential_nir2}) to the outer exponential LSB part of each
SBP. Column 5 gives the total apparent magnitude of the LSB component,
computed by extrapolating the fitted model (i.e. either an exponential
or a \med ) to $R^{\star}=\infty$.  Columns 6 through 9 list the radii
and magnitudes of the star-forming (P) and underlying stellar LSB
component (E), as obtained by profile decomposition. Following P96a,
we measure the respective radial extent ($P_{\rm iso}$, $E_{\rm iso}$)
and encircled magnitude ($m_{P_{\rm iso}}$,$m_{E_{\rm iso}}$) of each
component at an isophotal level $iso$, taken to be 23 \sbb\ for $J$
and 22 \sbb\ for $H$ and $K'$. The isophotal radii determined for the
sample BCDs at 23 $J$ \sbb\ turn out to be comparable to those
obtained from optical SBPs at 25 $B$
\sbb\ ($P_{25}$ and $E_{25}$ in P96a).
Column 10 lists the magnitude from an SBP integration out to 
the last data point, and total magnitudes from aperture measurements 
(cf. Sect. \ref{apphot_nir2}) are listed in column 11. The radii
$r_{\rm eff}$ and $r_{80}$, enclosing 50\% and 80\% of the SBP's
flux are included in column 12.
Finally, a {\em formal} \ser exponent for the whole SBP 
($\eta _{\rm SBP}$), for later comparison with literature data, 
is listed in column 13 of Table \ref{tab_phot_nir2}.
{ Column 14 gives the $J,H,K\arcmin $ calibration uncertainties for each
galaxy (Sect. \ref{cal_nir2}).}

\subsection{Colors of the underlying LSB host galaxy}
\label{lsbcolors_nir2}

Colors of the LSB host galaxy were derived as the error-weighted mean
of the color profiles, outside radii affected by SF activity
(cf. Sect. 3.5. in \cite{n03a}). Deviant points, being probably
affected by uncertainties in the sky determination, and local
residuals from the subtraction of background sources were rejected.
The mean colors of the host galaxies are shown at the right edge of
each color profile (Figs. \ref{fmkn314} - \ref{fn5058}).  { Table
\ref{tab_lsbcolors_nir2} lists the resulting colors, in the
photometric system described in Sect. \ref{cal_nir2}\,. $H-K'$ colors
given in column 3 were transformed to $H-K_s$ to facilitate
comparisons to other photometric systems. Colors in Table
\ref{tab_lsbcolors_nir2} can therefore be considered 2MASS colors. The
relatively large errors in columns 2 and 3 include conservative upper
limits to all uncertainties that would apply when comparing those
values to other colors in the 2MASS system. These are: (i) zeropoint
uncertainties (Table \ref {tab_phot_nir2}, column 14), (ii) the rms
scatter of the averaged color profile points, as upper limit to
effects of local background instabilities and fore-/background
sources, (iii) an estimate of the {\em systematic} uncertainties of
SBP derivation in the low $S/N$ regime (0.05, 0.07 and 0.1 mag for
good, medium and low $S/N$ data, respectively), (iv) an upper limit
(0.08 mag, cf. \cite{n03a}, Sect. 2.4.1) of possible uncertainties
from unknown color term transformations between our filters and those
used by 2MASS, (v) an additional uncertainty estimate of 0.05 mag in
$H-K_s$ to account for the $K\arcmin$ to $K_s$ transformation.}

\begin{table}
\caption{Colors of the host galaxy$^a$}
\label{tab_lsbcolors_nir2}
\begin{tabular}{lcc}\hline
Object          & $J-H$         &  $H-K_s$         \\
                & [mag]         & [mag]           \\ \hline
Mkn\,314        & 0.77$\pm$0.16$^b$& 0.26$\pm$0.18$^b$  \\ \hline 
Mkn\,209        & 0.37$\pm$0.13$^b$& 0.19$\pm$0.18$^b$   \\ \hline 
Mkn\,996	& 0.56$\pm$0.13	   & 0.37$\pm$0.16$^b$	\\ \hline   	
Mkn\,370	& 0.52$\pm$0.14	   & 0.16$\pm$0.17	\\ \hline   	
I\,Zw\,115	& 0.46$\pm$0.15$^b$& 0.31$\pm$0.26$^b$	\\ \hline   	
Mkn\,5  	& 0.54$\pm$0.15$^b$& 0.45$\pm$0.12$^b$	\\ \hline   	
Mkn\,600  	& 0.50$\pm$0.17$^{b,c}$& 0.08$\pm$0.29$^{b,c}$		\\ \hline   	
NGC\,6789  	& 0.42$\pm$0.12& 0.31$\pm$0.18$^{b}$		\\ \hline   	
Mkn\,324  	& 0.55$\pm$0.12& 0.22$\pm$0.18$^{b}$		\\ \hline   	
Mkn\,450  	& 0.51$\pm$0.16& 0.19$\pm$0.26$^{b}$		\\ \hline   	
NGC\,5058  	& 0.41$\pm$0.16& 0.24$\pm$0.25$^{b}$		\\ \hline   	

\end{tabular}\\[0.5em]
{Colors can be directly compared to data in the 2MASS photometric
system; for details of the photometric system and the various error
sources that are considered here, see Sect. \ref{lsbcolors_nir2} and
\ref{cal_nir2}.}\\
$^a$: Corrected for galactic extinction (see Sect. \ref{cal_nir2}).\\
$^b$: possible local instabilities in one SBP
at low $S/N$ levels.\\ 
$^c$: possible contamination by nebular emission
over a large portion of the LSB host galaxy\\
\end{table}

\subsection{Unsharp masking technique \& aperture photometry}
\label{apphot_nir2}

To improve on studies of fine morphological details, the {\em
hierarchical binning ({\rm hb}) transformation}, a modified unsharp
masking technique described in e.g. Papaderos (\cite{papaderosphd}),
was applied to the data. Features of interest are displayed in the 
insets of Figures \ref{fmkn314} -- \ref{fn5058}.

Total magnitudes of the galaxies were derived within polygonal apertures
which extend typically out to 1.5 Holmberg radii (column 11 of Table
\ref{tab_phot_nir2}), after removal of fore- and background sources from
the area of interest (cf. Sect. \ref{data_nir2}). These magnitudes can
generally be considered more accurate than those inferred from an
integration of SBPs (see \cite{n03a}).


\section{Results and discussion of individual objects \label{sample_galaxies_nir2}}
\subsection{Mkn 314 (NGC 7468, UGC 12329)}                                         
\label{mkn314}
\begin{figure*}[!ht]
\begin{picture}(18,10)
\put(0,0.1){{\psfig{figure=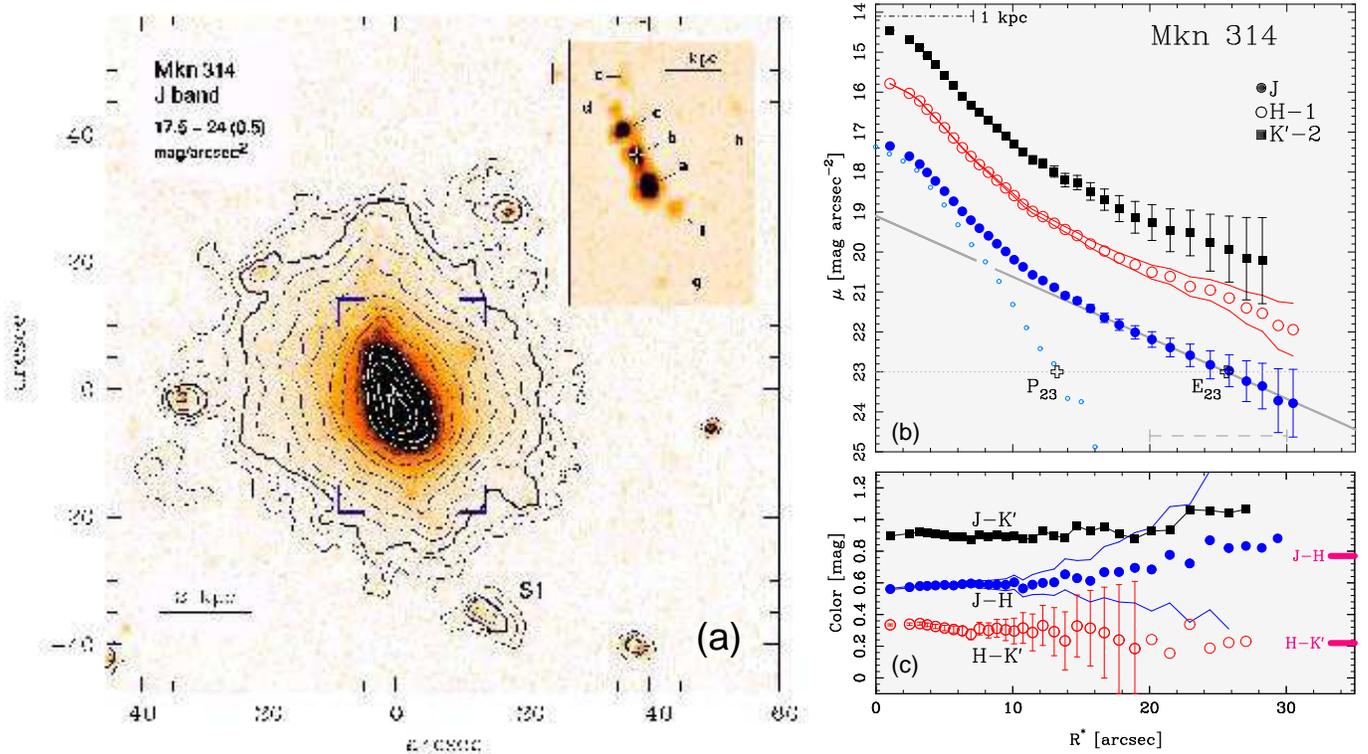,height=9.9cm,angle=0,clip=}}}
\put(10.9,4.01){{\psfig{figure=0221f1b.eps,width=7.1cm,angle=-90,clip=}}}
\put(10.95,0.19){{\psfig{figure=0221f1c.eps,width=7.1cm,angle=-90,clip=}}}
\put(9.2,1.5){\Large\sf (a)}
\put(11.8,4.3){\sf (b)}
\put(11.8,1.2){\sf (c)}
\end{picture}
\caption[]{{\bf (a)} 
Contours overlaid with a $J$ image of Mkn 314 ($D$=28.9
Mpc). Contours, corrected for Galactic extinction, go from 17.5 to
23.5 $J$ \sbb\ in steps of 0.5 mag. The 23\ $J$ \sbb\ isophote is
depicted by the thick contour.  The inset shows an unsharp-masked
(see Sect. \ref{apphot_nir2}) and magnified version of the central
region of the BCD (indicated by brackets in the contour map
image). Compact sources arranged in a bar-like structure along the
major axis of the BCD are marked, with the labeling of sources {\sf a}--{\sf c}
following the denomination by Mazzarella \& Boroson
(\cite{mazzarella93}).  The bright star-forming region {\sf b} is marked
with a white cross in both the main plot and the inset. The detached
SW source {\sf S1} shows active star formation.
{\bf (b)} Surface brightness profiles (SBPs) of Mkn 314 in the $J$, $H$
and $K_s$, corrected for galactic extinction.  For a better
visualization, the $H$ and $K_s$ SBPs are shifted by --1 and --2 mag,
respectively. The thick solid line shows an exponential fit to the
stellar LSB component in $J$ (cf. Sect. \ref{deco_nir2}), computed
within the radius range indicated by the long--dashed bar at the
bottom of the diagram.  The emission in excess to the fit (small open
circles) is attributable to the starburst component, which dominates
the light in the inner part of Mkn 314.  The isophotal radii $P_{23}$
and $E_{23}$ of the star-forming and LSB component at the surface
brightness level of 23 $J$ \sbb\ (horizontal dotted line) are
indicated.  The bar at the upper left indicates a galactocentric
radius of 1 kpc.
{\bf (c)} Color profiles, computed by subtraction of the SBPs (panel
{\bf b}). The thick lines at the rightmost part of the diagram
indicate the mean $J-H$ and $H-K'$ colors of the LSB component
(see Sect. \ref{lsbcolors_nir2}).}
\label{fmkn314}
\end{figure*}
%

As an iE-classified system, Mkn~314 represents the most common
morphological BCD type (LT86). Optical surface photometry as well as
deep H$\alpha$ imaging have been presented in C01ab. This
intrinsically luminous BCD (M$_{B}$=--18.5) was morphologically
selected as a candidate polar-ring galaxy (Whitmore et al. 1990, van
Driel et al. 2001), and was included in the study of Markarian
galaxies with multiple nuclei by Mazzarella \& Boroson
(\cite{mazzarella93}). Optical broad-band images by the latter authors
revealed three prominent maxima, roughly arranged along the major axis
of the galaxy (labeled {\sf a} through {\sf c} in Figure 1a).
Nordgren et al. (\cite{nordgren95}) found that regions {\sf a} and
{\sf c}, situated at comparable projected distances ($\approx$10\farcs
5 or 0.57 kpc and 9\farcs 4 or 0.52 kpc, respectively) from the
optically brightest, central source {\sf b}, show velocity differences
to the latter of 30$\pm$17 and 20$\pm$14 km s$^{-1}$. These values are
of the order of typical \ion{H}{I} velocity dispersions in BCDs (van
Zee et al. \cite{vanzee98}), and do not provide any strong evidence
for kinematical distortions which might indicate a previous merger
event. Narrow band images reveal that SF activity in the galaxy is
distributed along the northeast-southwest direction, in a bar-like
structure, which extends to about 5 kpc southwest from the nuclear
region (Deeg et al. \cite{deeg97}, C01b). BCDs showing such
bar-structures are relatively frequent (for instance Mkn~370, this
paper; Mkn~35, II~Zw~71, C01b; II~Zw~33, Walter et
al. \cite{walter97}) and are morphologically classified as ``chain
starburst'' in C01b.  The total H$\alpha$ luminosity of the galaxy
amounts to $\sim$6.1$\times$10$^{40}$\,erg\,s$^{-1}$
(\cite{cairos01b}, for the distance adopted here). In the H$\alpha$
frames, three strong SF regions are detected, aligned with the central
sources detected in the broad-band frames. The peak of the \ha\
emission is located at knot {\sf c}, whereas {\sf b} and {\sf a} are
moderate sources.

The morphologies in the optical and in the NIR basically coincide,
though the NIR frames provide a better spatial resolution. Smaller
condensations, surrounding the latter three major sources, were
identified (named {\sf d} -- {\sf h} in Fig. \ref{fmkn314}a ).

The starburst population is immersed in an extended older
population of stars, which displays elliptical isophotes and red colors
($B-R \approx$ 1, Cair\'os et al. 2001a, hereafter \cite{cairos01a};
C01b).  Since the chain of SF sources along the major axis of Mkn 314
is less prominent in the NIR than in optical wavelengths (cf. C01b),
slight differences in the outer slope of optical and NIR SBPs of the
LSB component are to be expected. In $J$ and $H$ we derive a scale
length of $\sim$1 kpc, somewhat smaller than the optical value
($\sim$1.2 kpc, C01a). NIR SBPs reveal a slight curvature for
\rr$\geq$20\arcsec\, pointing to a \flat\ SBP; however, the $S/N$
level in this outermost region is too low to corroborate this
conclusion.

A tail structure, formed by several SF knots, departs from the central
regions to the south-west, connecting with an extended
source named ``{\sf S1}'' in Fig. \ref{fmkn314}a\ . This object, which we
detect in all NIR bands, splits into two compact \ion{H}{ii} regions
on H$\alpha$ frames (C01b). The blue optical colors ($B-R\sim$0.5 mag;
C01b) of {\sf S1} are comparable to those observed in the central part
of the BCD (regions {\sf a} through {\sf c}, C01b). From the present data we cannot judge 
whether {\sf S1} is a gaseous or stellar interloper infalling into
Mkn 314: the projected velocity difference of 30 \kmsec\ (Nordgren et al. 1995)
to Mkn 314 is not unusual for close dwarf companions of BCDs
(Noeske et al. \cite{noeske01a}). Also, Taylor et al. (1994) find from
interferometric \ion{H}{I} observations of Mkn 314 little indication
for an interaction and describe merely an ``oval distortion'' of the
gaseous component.

\subsection{Mkn 209 (UGCA 281, I~Zw~36, Haro 29)}
\label{mkn209}
\begin{figure*}[!ht]
\begin{picture}(18,10)
\put(0,0.1){{\psfig{figure=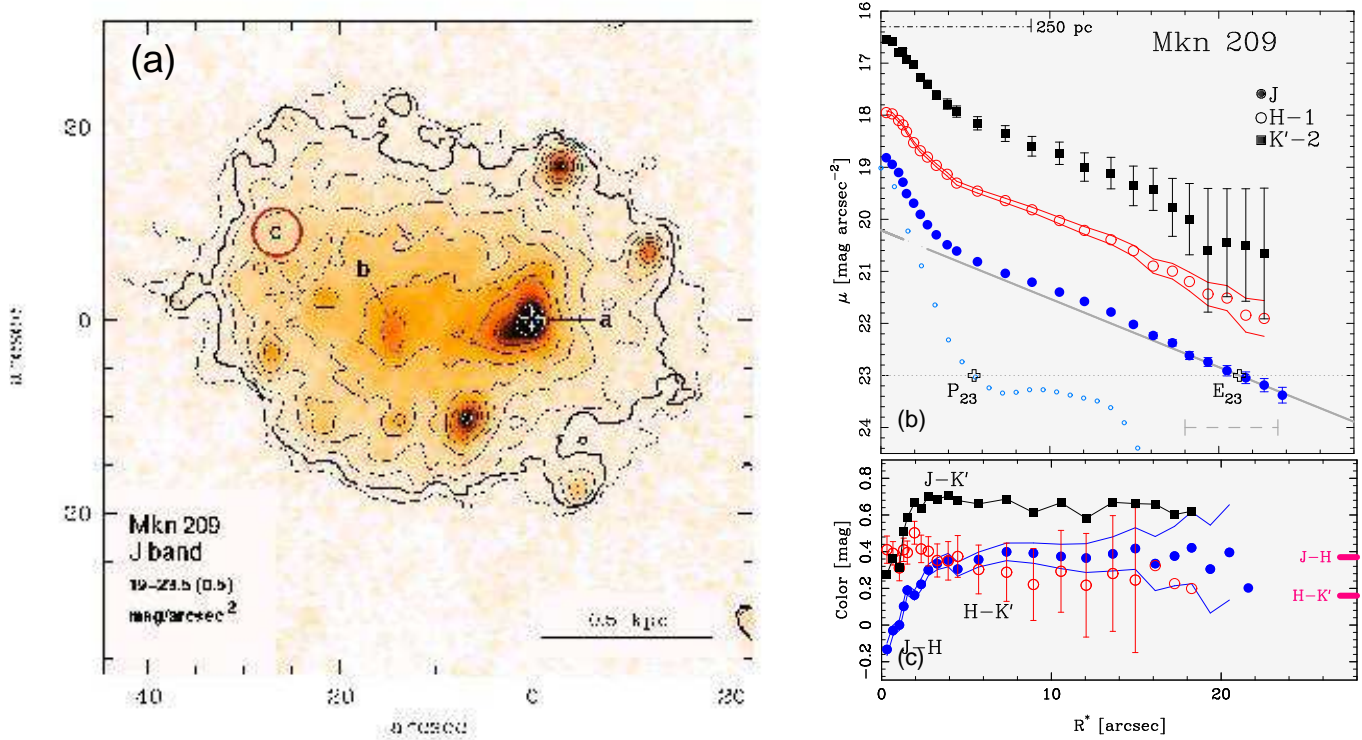,height=9.9cm,angle=0,clip=}}}
\put(10.9,3.92){{\psfig{figure=0221f2b.eps,width=7.0cm,angle=-90,clip=}}}
\put(10.95,0.19){{\psfig{figure=0221f2c.eps,width=7.05cm,angle=-90,clip=}}}
\put(1.6,9.0){\Large\sf (a)}
\put(11.8,4.28){\sf (b)}
\put(11.8,1.12){\sf (c)}
\end{picture}
\caption[]{Mkn 209 ($D$=5.8 Mpc). For explanations of symbols and 
labels, refer to Fig. \ref{fmkn314}. 
{\bf (a)} $J$ band image and isophotes. The \ha\ --emitting SF regions 
{\sf a} and {\sf b}, as well as the blue region {\sf c}, almost absent 
in NIR images, are marked. 
{\bf (b),(c)} Surface brightness and color profiles.} 
\label{fmkn209}
\end{figure*}

Mkn 209 provides an example of an intrinsically faint (M$_{B}$=
--14.2), compact (optical radius of
$\approx$ 0.6 kpc at 25 $B$ \sbb\ , P96a) and metal-deficient
(Z$\approx$\zsun/14, Izotov \& Thuan 1999) BCD with an iE-morphology
(LT86). This galaxy has been the subject of numerous studies,
due to its relative proximity ($D\approx 5.8$\,Mpc, Schulte-Ladbeck et
al. 2001).

The SF regions of Mkn 209 are morphologically reminiscent of those in
Mkn 5 (Sect. \ref{mkn5}), and are dominated by two bright sources
({\sf a} and {\sf b}, Fig. \ref{fmkn209}a).  Gil de Paz et al. (2003,
hereafter GMP03) found that the \ha\ emission peaks at the western SF
region {\sf a} (Fig. \ref{fmkn209}a). Additional diffuse \ha\ emission
is also present further to the east, close to region {\sf b} ($\approx
18$\arcsec\ from {\sf a}).  HST observations of knot {\sf a}
(Deharveng et al. 1994) showed the young stars to have ages $\leq$ 12
Myr, in agreement with spectral evolutionary synthesis models by
Mas-Hesse \& Kunth (\cite{mashesse99}), who derived a burst age of
$\approx$2.7 Myr.  Furthermore, spectroscopic studies of Mkn 209
revealed the presence of Wolf-Rayet stars (Schaerer et
al. \cite{schaerer99}, Guseva et al. \cite{guseva00}).

The source {\sf c} (Fig. \ref{fmkn209}a), detected in optical images
$\sim$15\arcsec\ northeast of {\sf b}, is nearly absent in the NIR. As
indicated by its blue colors on uncalibrated optical-NIR color maps,
and the presence of a local H$\alpha$ maximum (GMP03), region {\sf c}
may be strongly affected by nebular emission. The faintness of this
source in the NIR may indicate that the stellar population in this
area is still too young ($\la 8$\,Myr) to be dominated by Red Super
Giant (RSG) stars.

From an IUE UV spectral study, Fanelli et al. (1988) conjectured that
Mkn~209 could be a young galaxy, currently undergoing its first
episode of star formation. A later HST study by Deharveng et
al. (1994) showed, however, that the observed red colors could not be
attributed to RSG stars alone, but required the presence of an older
stellar population.  An old population was confirmed by
\cite{papaderos96a}, who reported almost constant, red ($B-R\sim 1$)
colors for the extended, elliptical stellar host galaxy of Mkn 209,
first detected by \cite{loose86}.  Recent $J$ and $H$ color magnitude
diagram studies of Mkn 209, using HST NICMOS data (Schulte-Ladbeck et
al. 2001), showed the presence of stars with ages $>$\,1 - 2 Gyr, in
agreement with the above results.

The SBPs of the LSB host can only be analyzed outside the very
extended plateau emission of the starburst, for \rr$>18$\arcsec . An
exponential fit, when extrapolated inwards, nowhere implies an
intensity brighter than the observed one. It leads, however, to a
local depression of the derived radial intensity distribution of the
starburst component at \rr$\sim 7$\arcsec\ . { In the $H$ band, an
exponential fit results in an unphysical decomposition solution
(starburst extent $P_{iso} \approx 0.02$\,kpc)}.  A \flat\ LSB
profile, fitted by a \med\ (Eq. \ref{med_nir2}) with a cutoff radius
$b\alpha\approx 14$\arcsec\ , could provide a more plausible
decomposition, but cannot be confirmed, due to the numerous fore- and
background sources which limit the reliability of the SBPs at their
outermost data points (\rr$\ga 20$\arcsec ). { We show a formal
decomposition using an exponential fit in the $J$ band only
(Fig. \ref{fmkn209}b and Table \ref{tab_phot_nir2}).}
\subsection{Mkn 996}
\label{mkn996}

\begin{figure*}[!ht]
\begin{picture}(18,10)
\put(0,0.1){{\psfig{figure=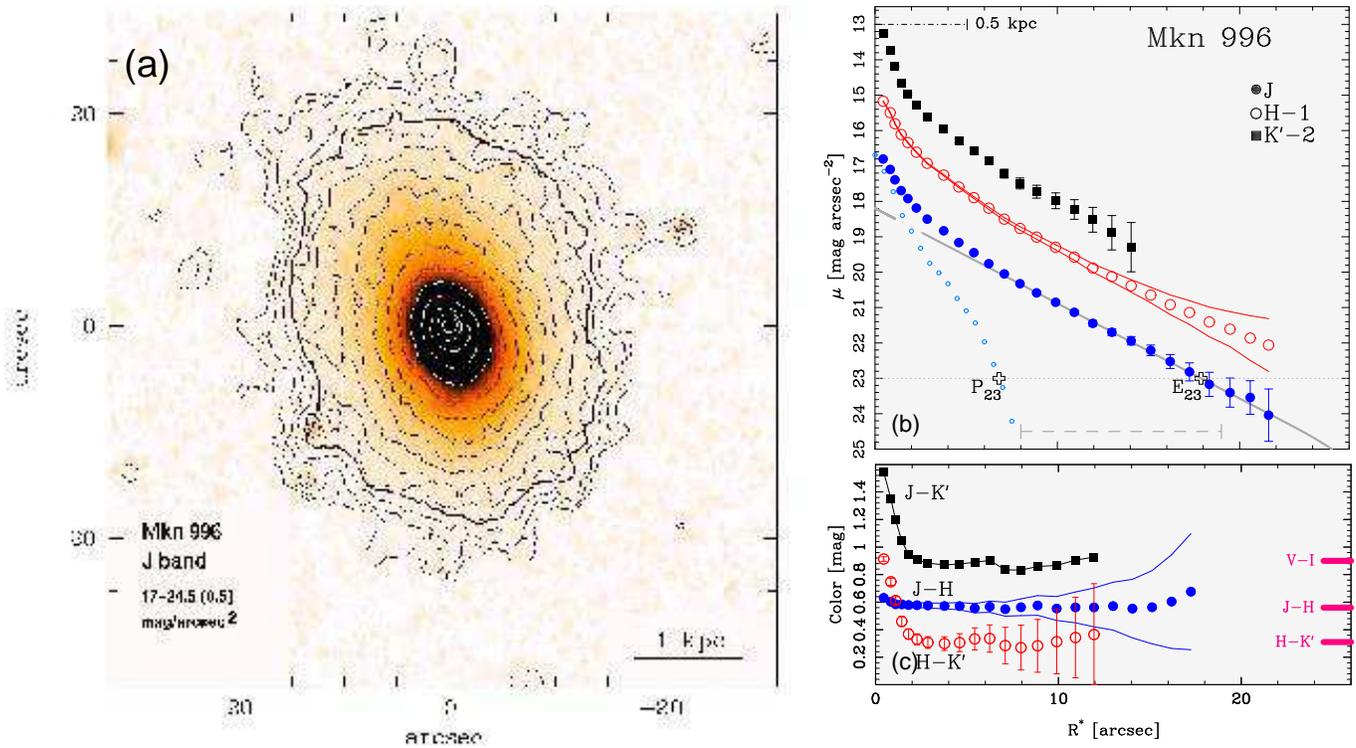,height=9.9cm,angle=0,clip=}}}
\put(10.9,3.92){{\psfig{figure=0221f3b.eps,width=7.0cm,angle=-90,clip=}}}
\put(10.95,0.19){{\psfig{figure=0221f3c.eps,width=7.05cm,angle=-90,clip=}}}
\put(1.6,9.0){\Large\sf (a)}
\put(11.8,4.28){\sf (b)}
\put(11.8,1.12){\sf (c)}
\end{picture}
\caption[]{Mkn 996 ($D$=20.4 Mpc). For explanations of symbols and labels, 
see Fig. \ref{fmkn314}.  {\bf (a)} $J$ band image and isophotes. 
{\bf (b),(c)} Surface brightness and color profiles. 
The indicated $V-I$ color of the LSB host galaxy has been adopted from
Thuan et al. (\cite{thuan96}).}
\label{fmkn996}
\end{figure*}

Mkn 996 ($M_{B}=-16.6$ mag) may be considered a prototypical example
of the nE BCD class.  Its starburst component is situated near the
geometrical center of its smooth LSB host galaxy and shows a very
compact morphology, with SF activity being confined to the inner
$\approx$3\arcsec\ (\rr$\approx$315 pc) of the BCD as evidenced by
Thuan et al. (\cite{thuan96}) using HST WFPC2 $V$ and $I$ data.
Optical spectra display strong Wolf-Rayet features, characteristic of
WN and WC stars, and imply a low oxygen abundance ($Z \approx$
\zsun/10), as well as an unusually high electron density of $\sim
5\times 10^4$ cm$^{-3}$ (Thuan et al. \cite{thuan96}, Schaerer et
al. \cite{schaerer99}).
The HST data by Thuan et al. (\cite{thuan96}) revealed dust patches
which extend to the north of the central SF region, as well as a
central two-arm spiral pattern with a size of 160 pc. 

The blue colors these authors derive in the vicinity of this central
region, $V-I\ga 0.15$, reflect the dominant emission of the young
stellar populations and ionized gas, superposed on the old, red LSB
component.  Interestingly, the NIR color profiles
(Fig. \ref{fmkn996}c) show within their central 2 arcseconds colors that
are significantly redder ($J-H\approx 0.6$, $H-K'\ga 0.8$) than those
expected for young starbursts.

Measurements in $4\arcsec\times 4\arcsec$ apertures, i.e. several
times larger than the angular resolution of the images, yield $(J-H)$
$\approx 0.61$, $(H-K')$ $\approx 0.7$ (after correction for the
underlying continuum), confirming that these colors are not due to
imperfect PSF-equalization during the generation of the color
profiles.  Such NIR colors cannot be accounted for by stellar and
nebular emission alone (see, e.g., Fig. 1 in Campbell \& Terlevich
\cite{campbell84}, or Fig. 1 in \cite{n03a}).  On the assumption of an
internal extinction of $E(H-K')\sim 0.3\dots 0.4$\,mag and a young
($\tau\sim 5$\,Myr), metal-poor ($Z \approx$ \zsun/10) starburst, such
colors could be reproduced.  Such an extinction would, however,
correspond to a large color excess in the optical, $E(V-I)\sim
1.9\dots 2.6$\,mag, in disagreement with the observed blue values of
$V-I$. We find that the $V$ and $J$ band intensity maxima coincide
within a fraction of an arcsecond, i.e. given the resolution and
alignment uncertainties, a significant shift is not found. Color maps
suggest increasing extinction only northwest of the region discussed
here.

Alternatively, the red $J-H$ starburst colors may be explained by the
presence of RSG stars. In metal-poor young stellar populations, the
fraction of such sources is higher than at solar metallicities,
leading to correspondingly redder $J-H$ values. This effect is not
properly reproduced by current stellar evolution models, which
therefore may predict too blue $J-H$ colors at low metallicities (cf.,
e.g., Sect. 4.2 in Vanzi \cite{vanzi03} and references
therein\footnote{Vanzi (\cite{vanzi03}) and Origlia et
al. (\cite{origlia99}) discuss the effect on $J-K$. However, the
colors of metal-poor RSGs ($J-H \sim 0.6 \dots 0.8$, $H-K \sim 0.1
\dots 0.2$, Elias et al. \cite{elias85} for RSGs in the SMC) imply
that an increased frequency of such sources will affect the integral
$J-H$ and $J-K$ colors of a young stellar population in a very similar
way.}).  If the burst is indeed already in its RSG-dominated phase
($\tau\ga 6.5$\,Myr), then the red $H-K'$ colors cannot be attributed
to dominant nebular emission any more.  Instead, these may originate
from blackbody emission of heated (few 100 $K$) dust, contributing
mainly to the $K$ band flux and thereby causing red $H-K'$ colors
(e.g. Campbell \& Terlevich \cite{campbell84}).

An intriguing property of this galaxy, discovered by Thuan et
al. (\cite{thuan96}), is that is has a population of several globular
clusters with an age of $\approx$ 10 Gyr in the very outskirts of the
LSB component, asymmetrically distributed towards the SW of the
galaxy, in contrast to the symmetric outer isophotes.

Surface photometry studies by Thuan et al. (\cite{thuan96}) showed
that for \rr$\ga$8\arcsec\ the stellar LSB host can be well fitted by
an exponential law with a scale length $\alpha$=420 pc.  The colors of
the outer regions of the galaxy ($V-I\approx$0.9) are characteristic
of an old LSB population.  These results are in good agreement with
our NIR surface photometry, from which we infer an exponential
intensity decrease of the LSB component with a scale length between
400 and 430 pc. Both our results, and those by Thuan et
al. (\cite{thuan96}) are in contrast with the conclusion by Doublier
et al. (1999), that Mkn 996 follows an overall de
Vaucouleurs-profile. This disagreement might stem from a subtle excess
for \rr $>18$\arcsec\ , visible in both our SBPs and those by Thuan et
al. (\cite{thuan96}), which at first glance points to a shallow, more
extended LSB population.  We believe, however, that this small excess
originates from the combined emission of the asymmetrically
distributed globular cluster population (see above).  These sources
cannot be properly resolved and subtracted using ground-based data,
and could therefore induce a slight flattening of the SBPs at large
radii.
We therefore fitted the SBPs for \rr$< 18$\arcsec\ in $J$ only,
i.e. at radii where the emission can be reliably attributed to the
unresolved stellar LSB population of Mkn 996.

\subsection{Mkn 370 (NGC 1036, UGC 02160)}
\label{mkn370}
\begin{figure*}[!ht]
\begin{picture}(18,10)
\put(0,0.1){{\psfig{figure=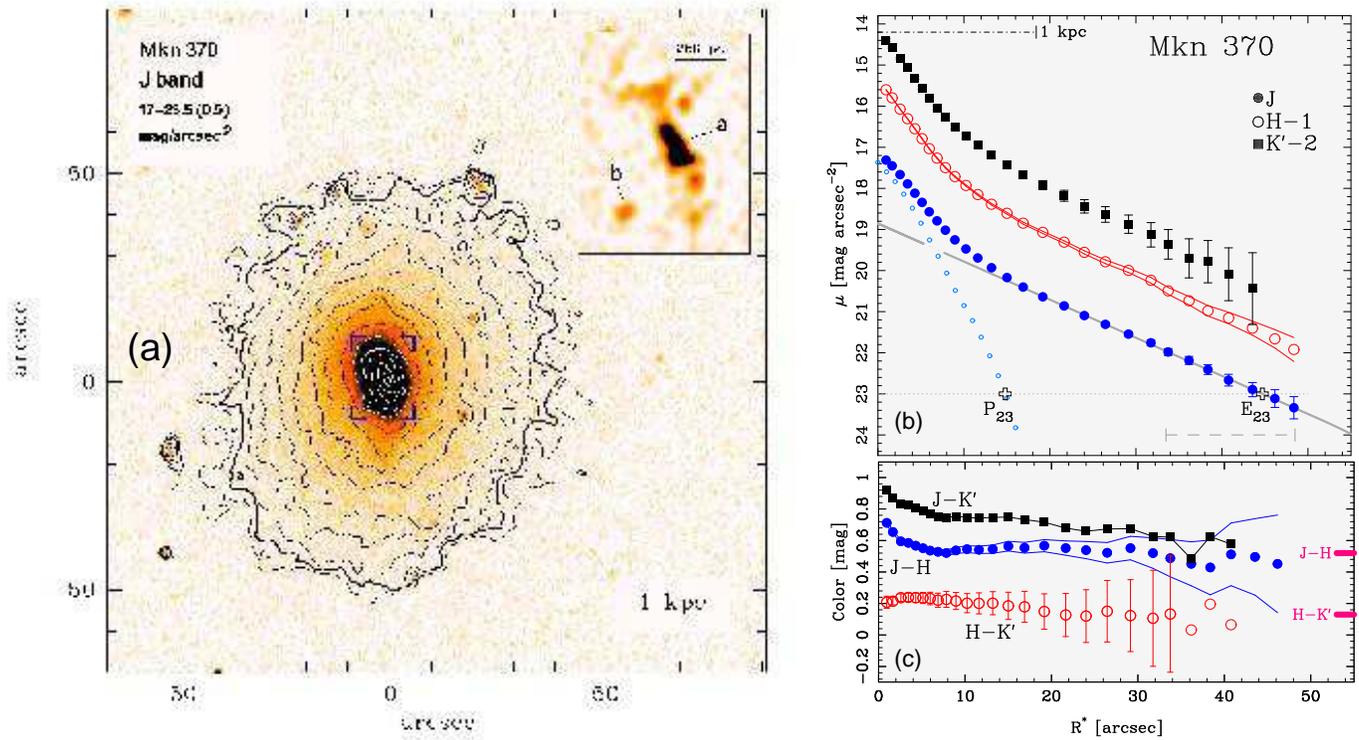,height=9.9cm,angle=0,clip=}}}
\put(10.9,3.92){{\psfig{figure=0221f4b.eps,width=7.0cm,angle=-90,clip=}}}
\put(10.95,0.19){{\psfig{figure=0221f4c.eps,width=7.05cm,angle=-90,clip=}}}
\put(1.6,5.2){\Large\sf (a)}
\put(11.8,4.28){\sf (b)}
\put(11.8,1.12){\sf (c)}
\end{picture}
\caption[]{Mkn 370 ($D$=11.2 Mpc). For explanations of symbols 
and labels, see Fig. \ref{fmkn314}. {\bf (a)} $J$ band image and isophotes.  
{\bf (b),(c)} Surface brightness and color profiles.}
\label{fmkn370}
\end{figure*}

Compared to other BCDs, the iE-classified (\cite{loose86}) BCD Mkn 370
is relatively metal-rich (Z$\approx$\zsun/2, M$_{B}$ = --17.1).
Deep broad-band surface photometry in the optical and high-resolution
color and H$\alpha$ maps were first presented in C01a,b\,, and a
dedicated spectrophotometric study of this BCD has recently been
published in Cair\'os et al. (\cite{cairos02}).  

Two major sources, separated by $\approx$7\arcsec\ (380 pc), are
located in the central region of the galaxy. These correspond to the
"double nucleus" catalogued in Mazzarella \& Boroson
(\cite{mazzarella93}) and are labeled {\sf a} and {\sf b} in Fig.
\ref{fmkn370}a, following the nomenclature of the latter authors.  As
apparent from the contrast-enhanced inset in the $J$ band image
(Fig. \ref{fmkn370}a), the elongated central region {\sf a} splits
into a brighter southwestern and a fainter northeastern source,
similar to what was reported by Nordgren et
al. (\cite{nordgren95}). The NIR images show several fainter sources,
roughly aligned with the major axis of region {\sf a}.

The positions of both {\sf a} and {\sf b} coincide in the optical and
in the NIR, although {\sf b}, which is a conspicuous source in the
blue, is only weakly detected in the NIR. The blue colors of this
source ($U-B=-0.79$), together with its emission line spectrum, flat continuum
and the large H$\alpha$ equivalent width ($EW($H$\alpha )$ $\approx$
500 \AA; Nordgren et al. 1995, Cair\'os et al.\cite{cairos02}) are
indicative of a young stellar population (ages $\leq$ 5 Myr) and
substantial ionized gas contribution. Knot {\sf a} presents slightly
redder colors ($U-B=$\,--0.60) and a high continuum level, with
absorption features that witness a substantial underlying population
of older stars.

The finding that knot {\sf b} shows no appreciable old stellar
background in NIR images, but is dominated by nebular emission,
indicates that the double-nucleus morphology of Mkn 370 is
attributable to extranuclear SF activity in knot {\sf b}.

The LSB population, traceable on our images out to \rr$\approx
50$\arcsec, provides about 80\% of the $J$ light.  Its NIR colors
(Table \ref{tab_lsbcolors_nir2}), as well as the integrated $B-J$
color of 2.3 mag, computed using data from Cair\'os et
al. (\cite{cairos02}), are both consistent with an old stellar
population. In NIR wavelengths the LSB component shows a roughly
exponential slope, with a possible, but not significant slight
flattening for \rr$\la 30$\arcsec. The $J$ band scale length,
$\alpha = 0.64$ kpc, is significantly smaller than the
$\alpha\sim$\,1\, kpc inferred from $B$ band data (Cair\'os et
al. \cite{cairos02}). This difference is most likely due to the
multiple extranuclear \ha -emitting sources, distributed roughly along
the optical major axis of Mkn 370 (Nordgren et al. 1995, Cair\'os et
al. \cite{cairos02}). Similar to what is discussed for Mkn 600
(Sect. \ref{mkn600}), these sources may contaminate the LSB emission
at large radii especially in optical wavelengths, artificially
increasing the scale length of the old population.

\subsection{I Zw 115 (UGC 09893)}
\label{izw115}
\begin{figure*}[!ht]
\begin{picture}(18,10)
\put(0,0.1){{\psfig{figure=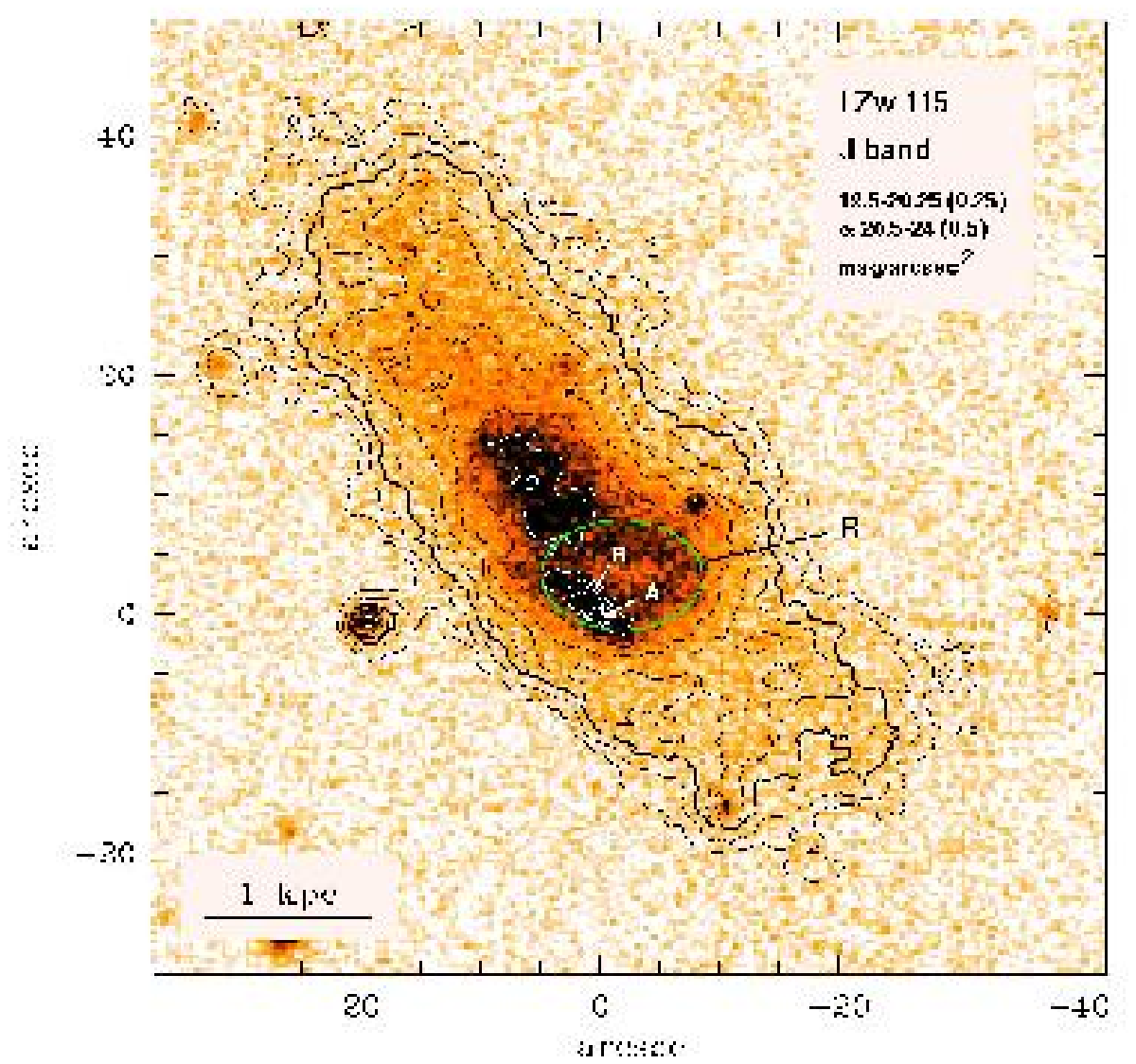,height=9.9cm,angle=0,clip=}}}
\put(10.9,3.92){{\psfig{figure=0221f5b.eps,width=7.0cm,angle=-90,clip=}}}
\put(10.95,0.19){{\psfig{figure=0221f5c.eps,width=7.05cm,angle=-90,clip=}}}
\put(1.6,5.2){\Large\sf (a)}
\put(11.8,4.28){\sf (b)}
\put(11.8,1.12){\sf (c)}
\end{picture}
\caption[]{I Zw 115 ($D$=15 Mpc). For explanations of symbols and labels, 
see Fig. \ref{fmkn314}.  {\bf (a):} $J$ band image and isophotes.  
{\bf (b),(c):} Surface brightness and color profiles. The thick grey
line shows a fit to the host galaxy using a modified exponential
distribution Eq. (\ref{med_nir2}) with $b,q=2.3,0.80$\ .}
\label{fizw115}
\end{figure*}

This iI -- classified (LT86) BCD (M$_{B}$=--16.4) was described in the
{\it Atlas of interacting galaxies} by Vorontsov-Vel'Yaminov
(\cite{vorontsov77}) as ``a pair of coalescents''.  Optical surface
photometry for this system was first presented in P96a, while further
$B$, $R$ and \ha\ images have recently been published in the catalogue of
Blue Compact Dwarf Galaxies by GMP03.

I~Zw~115 presents a peculiar morphology, displaying several intensity
maxima within an LSB component which, other than in the majority of
BCDs, shows boxy outer isophotes.  P96a conjectured that this could be
a signature of a dynamically unrelaxed underlying stellar component.
The derived $B-J$ color of the LSB host galaxy, $\sim 1.8\pm0.3$\,mag, is
consistent with an age of a few Gyr.
The brightest region, which spatially coincides in the optical and NIR
broad-band frames, is located towards the south-west of the LSB
component and is resolved into two smaller regions in the NIR images,
labeled {\sf A} and {\sf B} in Fig. \ref{fizw115}. The latter regions
seem to delineate, together with several other condensations, a
ring-like structure (``{\sf R}'' in Fig. \ref{fizw115}a) with a
projected diameter of $\sim$\,500\,pc in the southeast-northwest
direction. Northeast of this ring-like structure, several maxima are
detected, reminiscent of the tails of SF regions observed in
``cometary'' BCDs. However, H$\alpha$ emission in I Zw 115 is
primarily confined to one single SF region, which is displaced $\sim
15$\arcsec\ northeastwards with respect to {\sf a} and {\sf b} (see
the maps by GMP03).

The SBPs show an exponential decay at large radii, with a $J$ band
scale length of $\approx 530$\,pc, compatible to the one derived from
optical data (P96a, transformed to the distance adopted in this
paper). The \flat\ LSB profile detected by the latter authors is
confirmed by our analysis, which, however, implies a stronger central
depression $q$ and larger cutoff radius $b\alpha$ than those inferred
in the optical.  Using the approach described in
Sect. \ref{deco_nir2}, we can constrain the parameters of a \med\ to
$(b,q)\approx 2.4,0.80$.

\subsection{Mkn 5 (UGCA 130)}
\label{mkn5}
\begin{figure*}[!ht]
\begin{picture}(18,10)
\put(0,0.1){{\psfig{figure=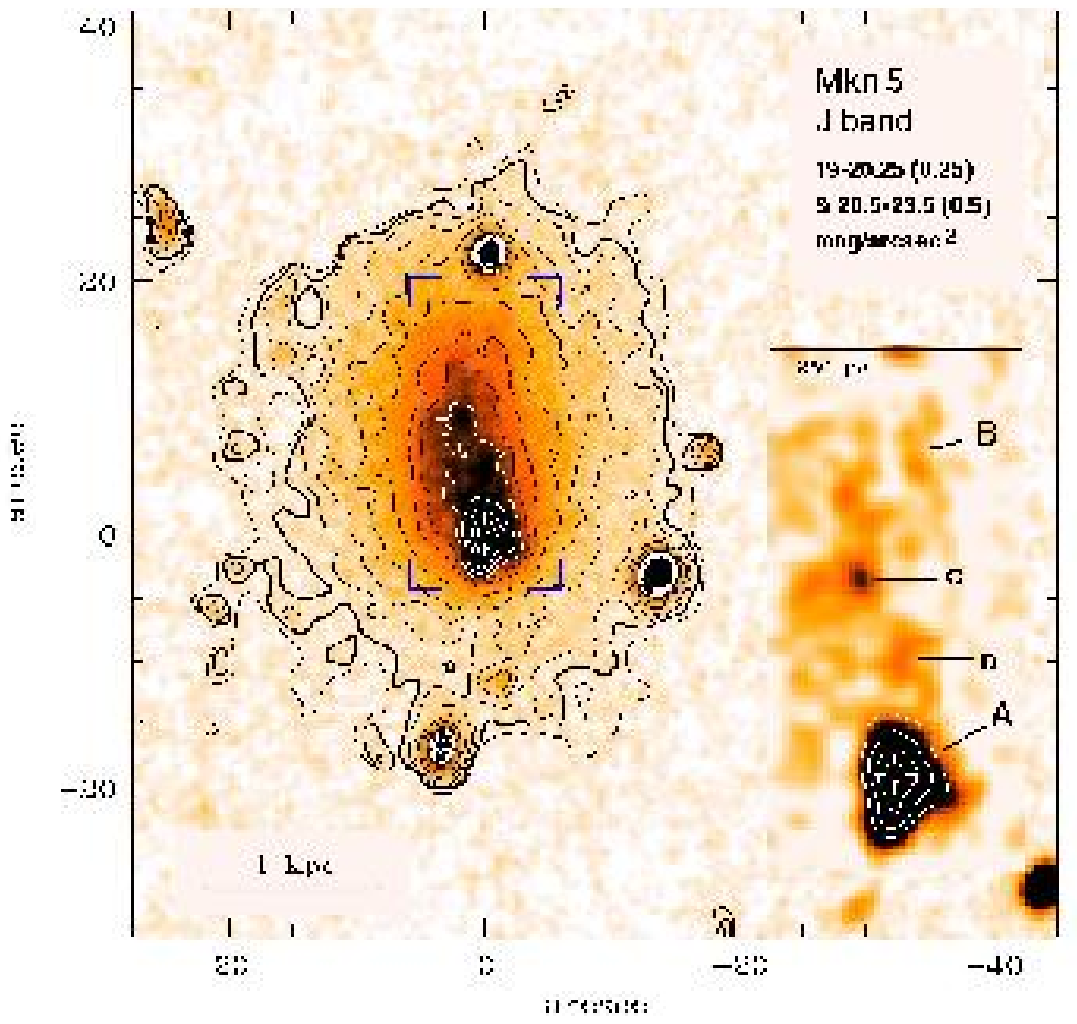,height=9.9cm,angle=0,clip=}}}
\put(10.9,3.92){{\psfig{figure=0221f6b.eps,width=7.0cm,angle=-90,clip=}}}
\put(10.95,0.19){{\psfig{figure=0221f6c.eps,width=7.05cm,angle=-90,clip=}}}
\put(1.6,9.0){\Large\sf (a)}
\put(11.8,4.28){\sf (b)}
\put(11.8,1.12){\sf (c)}
\end{picture}
\caption[]{Mkn 5 ($D$=15.3 Mpc). For explanations of symbols and labels, 
refer to Fig. \ref{fmkn314}. {\bf (a)} $J$ band image and isophotes.  
{\bf (b),(c)} Surface brightness and color profiles. The thick grey
line shows a fit to the LSB host galaxy using a modified exponential
distribution Eq. (\ref{med_nir2}) with $b,q=1.4,0.65$\ .}
\label{fmkn5}
\end{figure*}

While Mkn 5 can be considered an average BCD with respect to its
absolute magnitude ($M_B=-15.7$), oxygen abundance ($\sim$\zsun/7;
Izotov \& Thuan 1999) and the regular, elliptical isophotes of its LSB
host galaxy, it displays a peculiar, ``cometary'' morphology of its SF
regions. This latter classification (LT86) denotes the presence of a
dominant SF complex (``{\sf A}'' in Fig. \ref{fmkn5}a) towards one end
of an elongated stellar host galaxy, with fainter SF regions
distributed along the optical major axis. However, the ``iI,C''
BCDs, selected by this morphology of the SF regions, typically display
a more elongated, irregular host galaxy than that observed in Mkn 5
(see Noeske et al. \cite{noeske00}).

The dominant SF region {\sf A} shows WR features (Conti
\cite{conti91}) and is the locus of intense and moderately extended
nebular emission (EW(\ha)$\ga 60{\rm \AA}$; Noeske 1999).  The colors
of this knot ($U-B$=--0.78, $B-V$=0.49, $V-R$=0.18, $V-I$=0.19 within a
4\arcsec\ aperture) are compatible with a burst age $\leq$ 4 Myr
(Cair\'os 2000). Note that $B-V$ appears relatively red due to the
contribution of nebular line emission.

Among the SF regions forming the ``tail'' northwards of {\sf A},
source {\sf C} (Fig. \ref{fmkn5}a) is the brightest in the NIR,
in contrast to the optical images, where this knot is only marginally
detected. The second--brightest source in the optical, {\sf B},
situated $\sim 20$\arcsec\ north of {\sf A} , is barely seen even on
contrast-enhanced $J$ images. The \ha\ equivalent width at this
position is relatively low, $\sim$15 \AA, suggesting that its blue
colors are due to a young stellar population, rather than strong
nebular emission.

The NIR surface brightness profiles reveal an extended plateau feature
at high to intermediate intensity levels, similar to optical profiles
(C01a).
The exponential LSB profile, observed in the outskirts of Mkn 5,
flattens for small radii and can be well approximated by a \med\ with
$(b,q)=1.4,0.65$.  The $J$ band exponential scale length,
$\alpha\approx 5$\arcsec\ (0.37 kpc), is in good agreement with the
results by Noeske et al. (2001b) and C01a.

\subsection{Mkn 600}
\label{mkn600}
\begin{figure*}[!ht]
\begin{picture}(18,10)
\put(0,0.1){{\psfig{figure=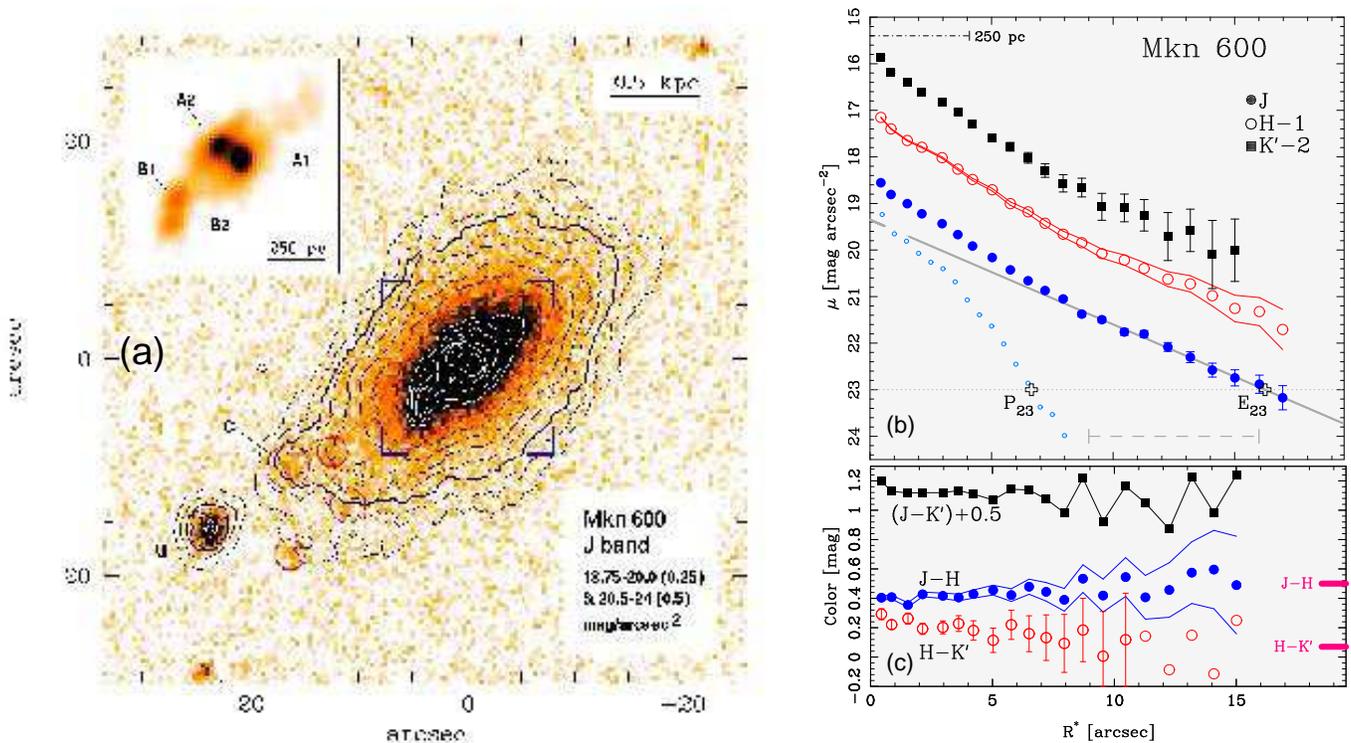,height=9.9cm,angle=0,clip=}}}
\put(10.9,3.92){{\psfig{figure=0221f7b.eps,width=7.0cm,angle=-90,clip=}}}
\put(10.95,0.19){{\psfig{figure=0221f7c.eps,width=7.05cm,angle=-90,clip=}}}
\put(1.6,5.2){\Large\sf (a)}
\put(11.8,4.28){\sf (b)}
\put(11.8,1.12){\sf (c)}
\end{picture}
\caption[]{Mkn 600 ($D$=12.6 Mpc). For explanations of symbols and labels, see 
Fig. \ref{fmkn314}. {\bf (a)} $J$ band image and isophotes. 
The circles at the SE of the galaxy's main body mark extranuclear knots, one of which 
(C) was studied in e.g. C01b. The
extended red object 'u', described by the same authors and probably
not belonging to Mkn 600, is marked.
{\bf (b),(c)} Surface brightness and color profiles.}
\label{fmkn600}
\end{figure*}

Star-forming activity in this iE BCD (M$_B$=--15.5, C01b;
$12+\log$\,O/H = 7.83, Izotov\& Thuan 1999, Cair\'os 2000) occurs
predominantly in its inner part (\rr$\leq$8\arcsec), where two
distinct maxima, {\sf A} and {\sf B}, are seen in optical images
(C01b). In the NIR images, each of the latter regions is resolved into
two condensations, respectively denoted {\sf A1, A2} and {\sf B1, B2}
in Fig. \ref{fmkn600}a. The ongoing SF activity in both {\sf A} and
{\sf B} is indicated by deep H$\alpha$ images (C01b) which show that
both regions coincide with local peaks in \ha\ line intensity and \ha\
equivalent width maps.  Both colors and \ha\ equivalent widths of {\sf
A} and {\sf B} point to burst ages $<$\,5 Myr (Cair\'os 2000). At
fainter surface brightness levels, optical and NIR data reveal a
regular LSB host galaxy which extends out to \rr$\sim$30\arcsec\, and
displays roughly elliptical isophotes. The colors of the latter
component ($V-R\approx 0.3$, $V-I\approx 0.8$, C01a; for NIR colors,
see Table \ref{tab_lsbcolors_nir2}) are indicative of a several Gyr
old stellar population.

A chain of fainter knots extends southeastwards from
the center of Mkn 600, and connects with a moderately bright, \ha
-emitting source (region {\sf c} in Fig. \ref{fmkn600}).  Source {\sf
c}, marginally detected in the NIR frames, emits strongly in H$\alpha$
($EW($H$\alpha$)$\approx 590{\rm \AA}$), and presents colors slightly
bluer than the central knots ($U-B=$ --0.64, --0.72 and --0.86 for
knots a, b and c, respectively), suggesting a possible propagation of
SF activity.

The hypothesis that the peculiar distribution of SF activity in Mkn
600 may be the result of an ongoing or recent interaction cannot be
discarded, given the presence of a nearby \ion{H}{I} companion (Taylor
et al. \cite{taylor93}). In our NIR images we detect no counterpart of
this object down to a surface brightness limit of $\approx$ 23
$J$\sbb.

As is evident from Fig. \ref{fmkn600}a, the influence of the
aforementioned extranuclear SF region {\sf c} is relatively small in
the NIR. This is also the case for other condensations in the vicinity
of {\sf c} (circles in Fig.\ref{fmkn600}a). Surface brightness
profiles, derived after subtraction of these sources, do therefore not
differ notably from those derived from the original images.
For \rr $>$9 arcsec, the SBPs were approximated by an exponential
distribution with scale lengths $\alpha\approx 4.8$\arcsec\ , slightly
smaller than those derived by C01a from optical data ($\alpha\approx
$\,5\farcs 5).  This difference is probably due to the stronger
contribution of region {\sf c} in optical bands (see also
Sect. \ref{mkn370}).

The integral colors we infer for Mkn 600 can hardly be reconciled with
those by Doublier et al. (\cite{doublier01}). These authors report an
integral $J-K$ color of $\sim$1 mag, much redder than the $J-K$ of
0.65 we derive.

\subsection{NGC 6789 (UGC 11425)}
\label{n6789}
\begin{figure*}[!ht]
\begin{picture}(18,10)
\put(0,0.1){{\psfig{figure=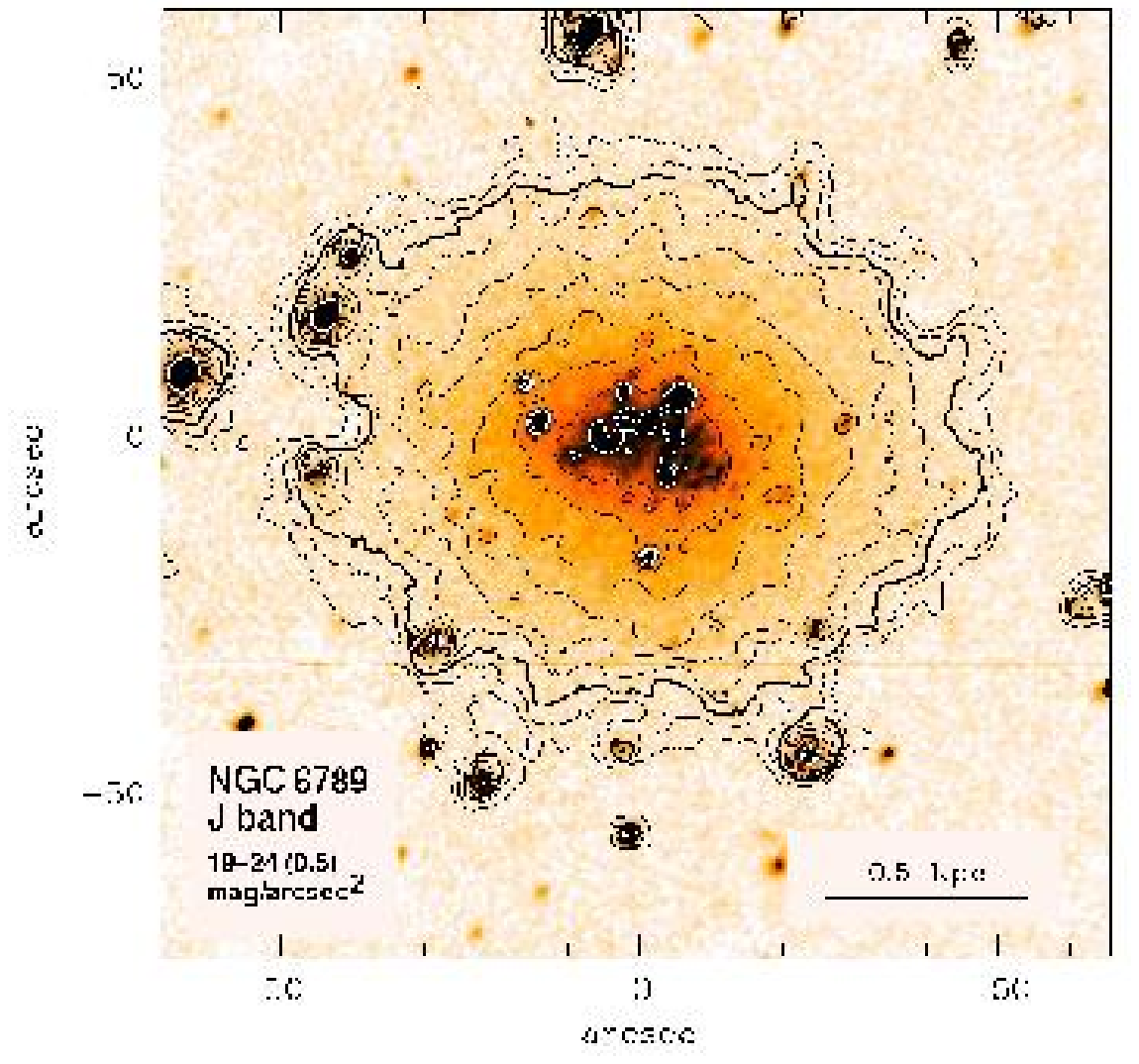,height=9.9cm,angle=0,clip=}}}
\put(10.9,3.92){{\psfig{figure=0221f8b.eps,width=7.0cm,angle=-90,clip=}}}
\put(10.95,0.19){{\psfig{figure=0221f8c.eps,width=7.05cm,angle=-90,clip=}}}
\put(1.6,9.0){\Large\sf (a)}
\put(11.8,4.28){\sf (b)}
\put(11.8,1.12){\sf (c)}
\end{picture}
\caption[]{NGC 6789 ($D$=3.6 Mpc). For explanations of symbols and labels, 
see Fig. \ref{fmkn314}.  {\bf (a):} $J$ band image and isophotes.  
{\bf (b),(c):} Surface brightness and color profiles. The thick grey
line shows a fit to the host galaxy using a modified exponential
distribution Eq. (\ref{med_nir2}) with $b,q=3.3,0.70$\ .}
\label{fn6789}
\end{figure*}

NGC~6789, an intrinsically faint (M$_{B}$=-14.3, Drozdovsky et
al. 2001) galaxy with an iE morphology, is located at a distance of
about 3.6 Mpc (Drozdovsky et al. \cite{drozdovsky01}) and therefore
belongs to the most nearby BCD candidates known to date, together with IC
10 (Richer et al. 2001) and IC 4662 (N03a).
Karachentsev \& Makarov (1998) found NGC 6789 to be very isolated,
situated in the local void.

Deep optical imaging (Drozdovsky \& Tikhonov \cite{drozdovsky00}) of
this system revealed properties typical among BCDs, in particular an
inner (\rr$\sim$150 pc), high surface-brightness ionizing stellar
population within a red ($V-I\approx$0.9 mag) underlying LSB host,
extending out to at least 0.6 kpc.

Subsequent HST/WFPC2 observations (Drozdovsky et al. \cite{drozdovsky01})
made it possible to resolve this galaxy into more than 15000 stars, and to
derive a minimum age of 1 Gyr from color-magnitude diagrams, in agreement
with the red $V-I$ LSB colors.

Inspection of the $J$ SBP slope shows that the LSB emission of NGC
6789 follows a \flat\ profile, which was fitted by a \med\ with a
cutoff radius of 3.3$\alpha$ and central depression of $q\approx 0.7$.
The scale length we derive in the outer regions, $\alpha\approx
200$\,pc (Table \ref{tab_phot_nir2}), differs from that inferred by
Drozdovsky \& Tikhonov (\cite{drozdovsky00}) ($\alpha\approx
280$\,pc). Since the moderately flattening \flat\ SBP of the LSB host
was not revealed by optical data, the latter authors
applied an exponential fit to the entire LSB emission (for
\rr$\ga 20$\arcsec ), thereby including slightly flattened parts of the
SBP inside the cutoff radius ($b\alpha \approx 38$\arcsec ).
The exponential part of the \med\ fitted to our $J$ band data has been adjusted
to the steeper, outer exponential regime of the profile (\rr$>39$\arcsec ).

The exponential scale lengths and extrapolated central surface
brightnesses inferred for NGC 6789, both by Drozdovsky \& Tikhonov
(\cite{drozdovsky00}) and in this paper (see Table
\ref{tab_phot_nir2}), indicate a compact structure of the
stellar host. This further supports the view that NGC 6789 displays
properties typical of BCDs, rather than those of more diffuse dwarf
irregulars.

\subsection{Mkn 324 (UGCA 439)}
\label{mkn324}
\begin{figure*}[!ht]
\begin{picture}(18,10)
\put(0,0.1){{\psfig{figure=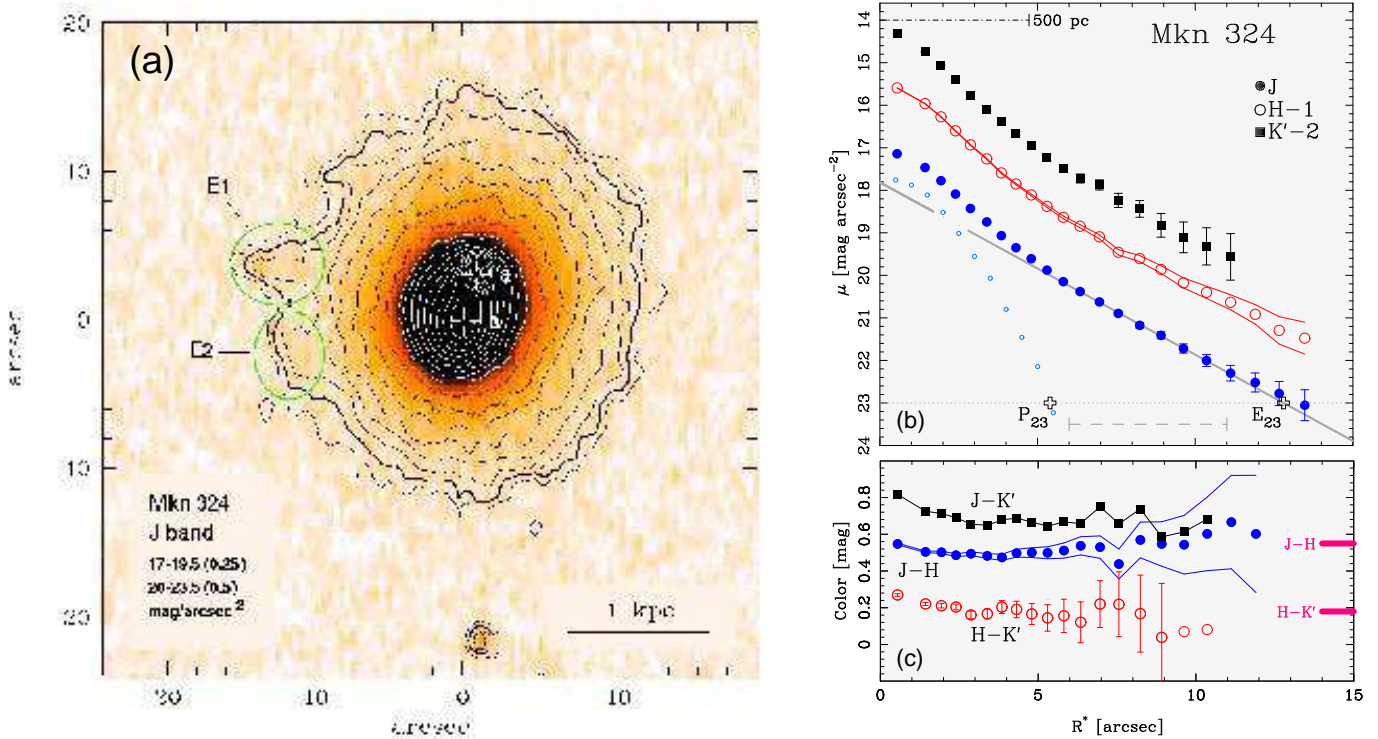,height=9.9cm,angle=0,clip=}}}
\put(10.9,3.92){{\psfig{figure=0221f9b.eps,width=7.0cm,angle=-90,clip=}}}
\put(10.95,0.19){{\psfig{figure=0221f9c.eps,width=7.05cm,angle=-90,clip=}}}
\put(1.6,9.0){\Large\sf (a)}
\put(11.8,4.28){\sf (b)}
\put(11.8,1.12){\sf (c)}
\end{picture}
\caption[]{Mkn 324 ($D$=21.8 Mpc). For explanations of symbols and 
labels, refer to Fig. \ref{fmkn314}.  {\bf (a)} $J$ band image and 
isophotes. The star--forming knots {\sf a} and {\sf b} 
(Mazzarella \& Boroson \cite{mazzarella93}) are labeled. 
{\sf E1} and {\sf E2} mark the diffuse sources adjacent to the east
of the LSB component (see text).
{\bf (b),(c)} Surface brightness and color profiles.}
\label{fmkn324}
\end{figure*}

Mkn~324 (M$_{B}$=--16.5) is a compact iE BCD which displays two
distinct SF regions close to its geometrical center, referred to by
Mazzarella \& Boroson (\cite{mazzarella93}) as {\sf a} and {\sf b}
(Fig. \ref{fmkn324}a) and studied by the same authors.

Surface photometry in the optical has been previously presented by
Doublier et al. (\cite{doublier97},\cite{doublier99}) and
\cite{cairos01a}. The latter authors (\cite{cairos01b}) also computed
optical color maps, and disclosed in deep \ha\ images a complex
morphology of the ionized gas, most notably a large supershell
expanding to the northwest of the starburst region.  The \ha\
intensity was found to peak in between the pair of bright SF regions
{\sf a} and {\sf b} (\cite{cairos01b}, Petrosian et
al. \cite{petrosian02}).  In its LSB periphery, Mkn~324 is delimited
by regular outer isophotes, and shows no conspicuous signatures of a
strong gravitative perturbation.

It appears nevertheless possible that the starburst activity of this
system has been triggered by a distant low-mass companion.
High-resolution \ion{H}{I} VLA maps (van Zee et al. \cite{vanzee01})
have recently unveiled a gas cloud with a mass of
$\sim 2\times10^8$\,\msun, approximately 115\,kpc NW of the BCD (values
transformed to the distance adopted in this paper), with a projected
velocity difference of $\sim 100$\,km\,s$^{-1}$ to Mkn 324. A follow-up
analysis of this system by Cair\'os et al. (2004, in prep.) has revealed an
intrinsically faint, low-surface-brightness optical counterpart to this
\ion{H}{I} source. In this respect, it appears noteworthy that the 
\ion{H}{I} component of Mkn 324 shows an overall solid-body rotation, 
though with conspicuous kinematical peculiarities, reminiscent of
tidal tails (van Zee et al. \cite{vanzee01}).  2D studies of the
ionized gas component of the BCD (Petrosian et al. \cite{petrosian02})
have shown a largely chaotic velocity field, with some underlying
regular pattern.

Our surface photometry reveals in all NIR bands an exponential LSB
distribution in the radius range 5\farcs9$\la$\rr$\la$11\arcsec, with
a moderately strong starburst emission.  For this outer LSB component,
we derive an exponential $J$ band scale length of $\approx 280$\,pc,
in agreement with the $I$ band scale length derived by C01a.  Some
excess emission above the exponential SBPs for \rr\ $>$11\arcsec\ is
presumably due to imperfect subtraction of diffuse sources adjacent to
the eastern edge of the LSB component (sources {\sf E1} and {\sf E2}
in Fig. \ref{fmkn324}a).

The possibility that this excess above the exponential fit
at large radii is a signature of a \ser\ law with an exponent
$\eta\ga$\,1\, could be dismissed. \ser\
fits over different radii in the interval
5\farcs9$\leq$\rr$\leq$15\arcsec\,, yielding exponents $\eta$ between 1.4
and 2, could not be extrapolated inwards without exceeding the observed
SBPs at small radii, i.e., they fall short of describing the
SBP of the LSB component.

\subsection{Mkn 450 (UGC 08323)}
\label{mkn450}
\begin{figure*}[!ht]
\begin{picture}(18,10)
\put(0,0.1){{\psfig{figure=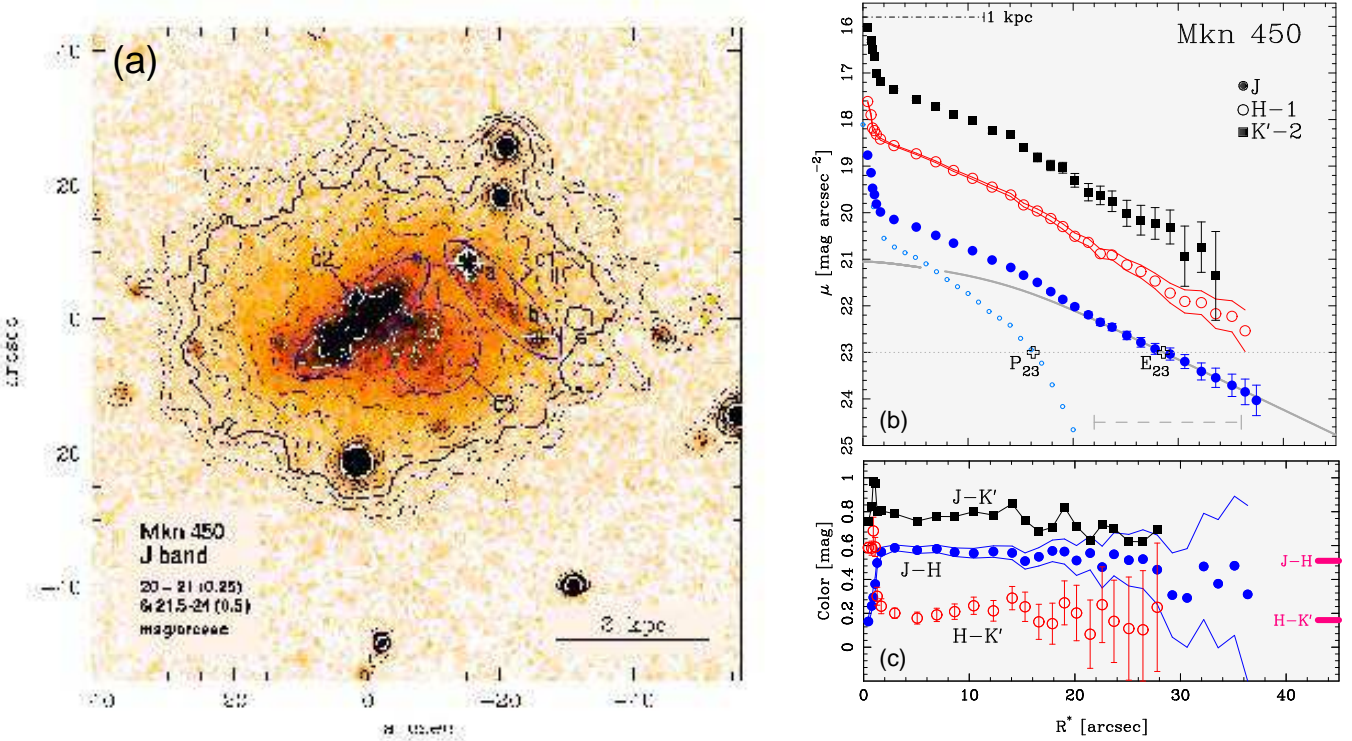,height=9.9cm,angle=0,clip=}}}
\put(10.9,3.92){{\psfig{figure=0221f10b.eps,width=7.0cm,angle=-90,clip=}}}
\put(10.95,0.19){{\psfig{figure=0221f10c.eps,width=7.05cm,angle=-90,clip=}}}
\put(1.6,9.0){\Large\sf (a)}
\put(11.8,4.28){\sf (b)}
\put(11.8,1.12){\sf (c)}
\end{picture}
\caption[]{Mkn 450 ($D$=17.9 Mpc). For explanations of symbols and labels, refer to
Fig. \ref{fmkn314}.  {\bf (a)} $J$ band image and isophotes. The bright
star--forming knots {\sf a} and {\sf b} are labeled. {\sf c1}\dots{\sf
c3} mark distinct star-forming complexes described in the text.  {\bf
(b),(c)} Surface brightness and color profiles. The thick grey line
shows a fit to the host galaxy using a modified exponential
distribution Eq. (\ref{med_nir2}) with $b,q=1.5,0.65$\ .}
\label{fmkn450}
\end{figure*}

Star-forming regions within this iE BCD are distributed over a large
surface fraction of its elliptical stellar LSB host. Individual
condensations appear to be arranged within 3 distinct associations.
Close to the geometrical center of the LSB host, an elongated
structure ({\sf c2} in Fig. \ref{fmkn450}) with an extent of roughly 2
kpc is seen along the southeast-northwest direction. It is displaced
relative to the optical major axis of the galaxy, which roughly runs
in the east-west direction. Adjacent to the southwestern edge of {\sf c2},
a fainter complex of knots, {\sf c3}, can be discerned.  A third strip
of SF knots, {\sf c1}, is seen NW of regions {\sf c2} and
{\sf c3}, oriented roughly perpendicular to the former.  The latter
feature hosts the brightest individual condensations, {\sf a} and {\sf
b} in Fig. \ref{fmkn450}a. The photometric properties of each source
({\sf a:} $m_J\sim 17$\,mag, $J-H\sim 0.3$\,mag, $H-K'\sim 0.6$\,mag;
{\sf b:} $m_J\sim 18$\,mag, $J-H\sim 0.2$\,mag, $H-K'\sim 0.6$\,mag),
derived after correction for the surrounding continuum, point to young
stellar populations and strong contributions of ionized gas. Recent
\ha\ maps by GMP03 confirm that SF activity in Mkn 450 is largely
confined to regions {\sf a} and {\sf b}, while only faint \ha\
emission is present at the location of the features {\sf c2} and {\sf
c3}.

The alignment of the morphological complex {\sf c1} with respect to
the LSB host is atypical among iE BCDs; usually, such elongated
sequences of SF regions are found closer to the centers of such
objects, at position angles parallel rather than perpendicular to the
major axis of the LSB host (see e.g. the morphological catalogs by
Telles et al. \cite{telles97b}, C01a, GMP03).  Mkn 450 is a
field BCD (Popescu et al. \cite{popescu99}) which shows no
peculiarities among its other properties, such as its $B$ magnitude of
M$_B$ = --16.70 (Vennik et al. \cite{vennik00}), and both a moderate
metal-deficiency (1/5 $Z_{\odot}$, Garnett \cite{garnett90}) and star
formation rate ($\sim 0.1$\,\msun\,yr$^{-1}$; Popescu et
al. \cite{popescu99}, value transformed to the distance adopted in
this paper).

Subtraction of only the brightest irregular SF regions
reveals a modest central flattening of the stellar LSB host galaxy,
well described by a \med\ with ($b$,$q$)=(1.5,0.65). 

\subsection{NGC 5058 (UGC 08345, MKN 786)}
\label{n5058}
\begin{figure*}[!ht]
\begin{picture}(18,10)
\put(0,0.1){{\psfig{figure=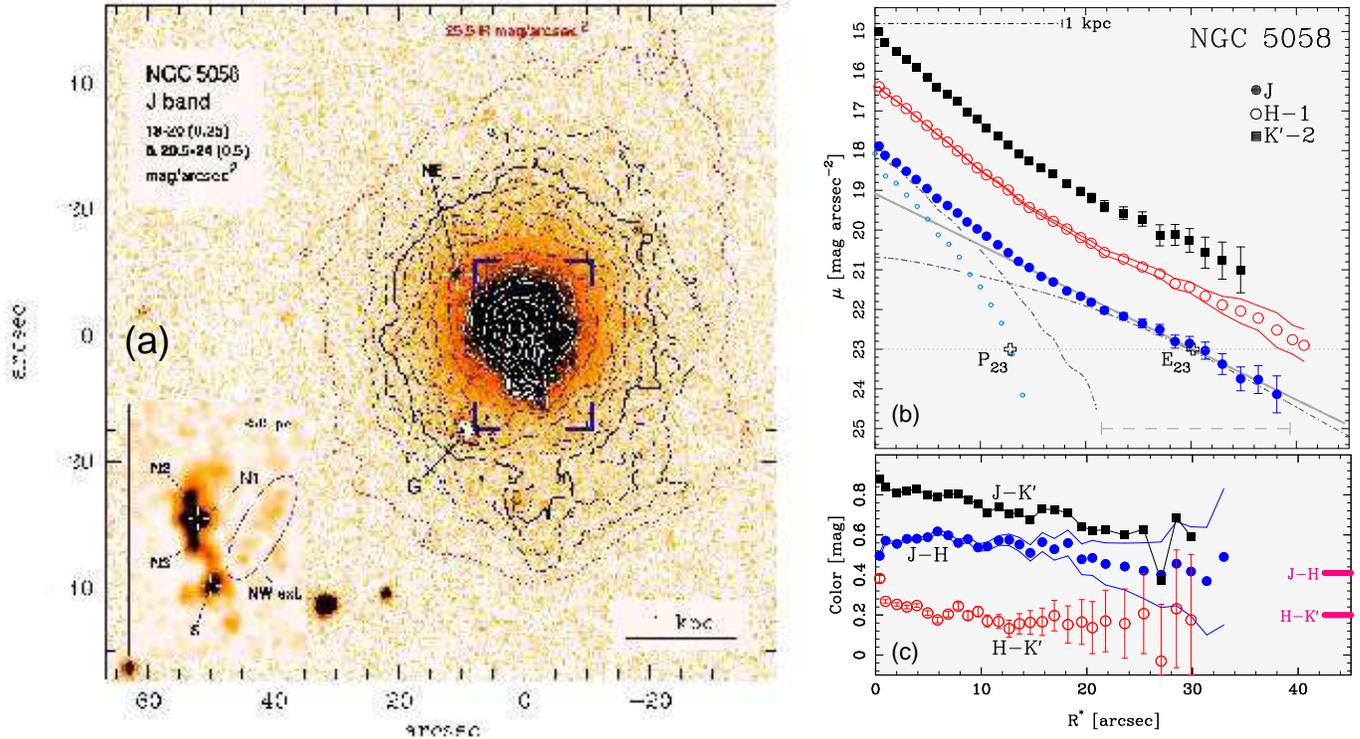,height=9.9cm,angle=0,clip=}}}
\put(10.9,3.92){{\psfig{figure=0221f11b.eps,width=7.0cm,angle=-90,clip=}}}
\put(10.95,0.19){{\psfig{figure=0221f11c.eps,width=7.05cm,angle=-90,clip=}}}
\put(1.6,5.2){\Large\sf (a)}
\put(11.8,4.28){\sf (b)}
\put(11.8,1.12){\sf (c)}
\end{picture}
\caption[]{NGC 5058 ($D$=11.6 Mpc). For explanations of symbols and labels, refer to
Fig. \ref{fmkn314}.  {\bf (a)} $J$ band image and isophotes.  The
blue north-eastern knot (NE) and the red source to the southeast (G) are
marked. Small white crosses mark the northern (N1) and southern (S)
maxima. The inset shows a magnified, unsharp-masked image of the
region bracketed in the main image. See the text.
{\bf (b),(c):} Surface brightness and color profiles. The thick solid
line illustrates an exponential fit to the stellar LSB component in
$J$ (cf. Sect. \ref{deco_nir2}), computed in the radius range
indicated by the light gray, long--dashed bar at the bottom of the
figure.  The emission in excess to this fit is shown by the small open
circles.  The thin dash--dotted curve gives an example of alternative
decomposition solutions, using a type V model for the LSB component
(Eq. \ref{med_nir2} with $b,q=3.8,0.9$) . The residuals to this fit are
also plotted as a dash--dotted line.
\label{fn5058}}
\end{figure*}

Deep images of this galaxy show two maxima (Fig. \ref{fn5058}a)
within a common, extended LSB component, which at low surface
brightness levels displays an asymmetric, curved morphology (see the
25.5 $R$ \sbb\ isophote in Fig. \ref{fn5058}a).

NGC 5058 was previously classified as a galaxy pair (KPG 370; Karachentsev
\cite{Karachentsev72}). The components of this pair (N \& S, Tifft 1982)
are most likely identical to the maxima visible in Fig. \ref{fn5058}a,
as can be inferred from their relative positions (projected separation
8\farcs 7 at a position angle 12\degr; Tifft 1982 for components N \&
S). We therefore adopted the latter denomination (cf. the inset in
Fig. \ref{fn5058}a).

On contrast-enhanced NIR images, source {\sf N} resolves into 3
distinct regions ({\sf N1, N2, N3} in Fig. \ref{fn5058}a) . Fainter
extensions emanate from this central source to the north, and
southwards, connecting with source {\sf S}. A third extension (``{\sf
NW}'' in Fig. \ref{fn5058}a) extends northwestwards from the southern
source {\sf S}.  New \ha\ images by GMP03 show that the SF activity
peaks close to source {\sf S}, and around a northeastern component,
detached from the central regions (``{\sf NE}'' in
Fig. \ref{fn5058}). The same data reveal only weak \ha\ emission along
the {\sf NW} extension, suggesting that this feature is not mainly due
to nebular emission.

Since NGC 5058 is located at a small angular separation (11\degr .6)
from the center of the Virgo Cluster, redshift-based distance
determinations using Virgocentric infall models suffer from the Triple
Value Problem (Teerikorpi et al. \cite{teerikorpi92}). An alternative
distance determination by the same authors, based on the Tully-Fisher
relation, yields a large distance of 37.7 Mpc, even outside
the range of values allowed within the uncertainties of the
Triple Value Problem (See Fig. 1 in Teerikorpi et
al. \cite{teerikorpi92}). We suggest that the \ion{H}{i} velocity width 
that these authors use, 118\,km\,s$^{-1}$, may be partly enlarged 
by the southern component {\sf S}, which, by its velocity difference
to component {\sf N} (58\,km\,s$^{-1}$, Tifft \cite{tifft82}), might
be a kinematically distinct subunit of NGC 5058. 

Dynamical distortions or even a past merger event in this galaxy
appear also possible in view of the peculiar shape of the isophotes in
the LSB regime, as mentioned above.  Irregularities south of the
center of the galaxy are traceable down to the limits of the $J$ band
data ($\sim$\,24\,$J$\sbb ). These condensations in the southernmost
outskirts of the galaxy show no obvious \ha\ emission (GMP03).

We adopt here a distance $D=11.6$\,Mpc, using the velocity of NGC 5058
with respect to the center of the Virgo Cluster (see \cite{n03a}) and
$H_0=75$\,km\,s$^{-1}$\,Mpc$^{-1}$; note that also for much larger 
distances, up to $D\la 31.1$\,Mpc, the resulting absolute magnitude would 
be $>-18B$\,mag, still qualifying NGC 5058 as a BCD.

In view of the complex morphology at all surface brightness levels,
and of wide-spread SF activity and further condensations, a
decomposition of the surface brightness profiles is rendered
problematic.  For \rr$>21.6$\arcsec\, i.e. outside strong
irregular emission in the $J$ images, the SBPs can be approximated by
an exponential, albeit with some systematic differences
(Fig. \ref{fn5058}b). The SBPs suggest however the presence of a \flat\
SBP at large radii. An example decomposition by means of a \med\ is
shown in Fig. \ref{fn5058}b. The parameters of this function cannot be
reliably constrained, given the extended influence of irregular and
starburst emission, and the low $S/N$ levels ($\mu _J\sim
22\dots23$\sbb ) at which we suspect the flattening occurs. The
results of the profile decomposition given in Table
\ref{tab_phot_nir2} therefore refer to the exponential fit to the LSB
host.
\section{Summary}
\label{summary}

This work is part of a series of papers that describe an imaging
study of a large sample of Blue Compact Dwarf (BCD) galaxies in the
NIR.  We have presented deep $J$, $H$ and $K'$ NIR images for 11
northern BCDs. These data, together with those presented in N03a and
C03a, constitute the largest sample of BCDs analyzed so far in the
NIR. Hitherto unreached deep surface brightness limits (23 $\dots$ 24
$J$\sbu\ ) allow detailed surface photometry of the evolved,
extended stellar host galaxies of such objects, to derive the
structural parameters of their light distributions and their colors.

Radial surface brightness profiles (SBPs) were decomposed into the
contributions of the extended stellar host galaxy, and of the
starburst component, the latter encompassing the emission of current
star-forming activity and recently formed young stars.  This task was
accomplished through the extrapolation of fits to the SBPs of the host
galaxy, at larger radii where the emission of the centrally
concentrated starburst component vanishes. In the NIR, the fractional
contribution of the starburst is lower than in the optical, making it
possible to study the host galaxies closer to their center, and to
test for central deviations from the exponential distributions that
the host galaxy SBPs typically display in their outer parts.

For each object we present $J$ contour maps, overlaid with the $J$
images and the derived surface brightness and color profiles. A
detailed description of the morphology and peculiar properties of each
galaxy are also provided.

The results of this work support the general findings described for
BCDs in the NIR by N03a, and can be summarized as follows:

\begin{enumerate}

\item At larger galactocentric radii, the NIR SBPs of the evolved
stellar host galaxies show a roughly exponential decay, in agreement
with previous results from optical studies.  Exponential scale lengths
derived in the optical and NIR are mostly compatible within the uncertainties, 
indicating that old stellar components of BCDs typically have no appreciable
color gradients at intermediate and large radii.

\item In 4 out of 11 sample BCDs, the SBP of the host galaxy shows a
central flattening with respect to its exponential decay at larger
radii, typically within \rr $\ga$ 2 -- 3 exponential scale lengths of
the outer profile. In 4 additional objects, such inwards-flattening
exponential SBPs (denoted ``type V'' in Binggeli \& Cameron
\cite{binggeli91}) of the stellar host galaxies are found to be
likely. These results support our previous notion (\cite{n03a}) that
centrally flattening SBPs may be frequent in stellar hosts of BCDs,
but largely remain undetected at optical wavelengths where central
starburst emission is more dominant.

\item
The fact that NIR data are less affected by nebular emission and dust
extinction than optical data, yet more susceptible to emission of hot
dust, allows a more profound interpretation of the morphology and
properties of individual sources of the star-forming regions. Such a
less biased view of the spatially resolved star formation history of
BCDs may be essential for understanding the probably complex mechanisms
that govern their massive, spatially extended star-forming activity.

\end{enumerate}


\begin{acknowledgement}
Research by K.G.N. has been supported by the Deutsche
Forschungsgemeinschaft (DFG) grants FR325/50-1 and FR325/50-2.
P.P. and K.J.F. received support from the Deutsches Zentrum f\"ur
Luft-- und Raumfahrt e.V. (DLR) under grant 50 OR 9907
7. L.M.C. acknowledges support from the European Community Marie Curie
Grant HPMF--CT--2000--00774. The authors wish to thank the staff of
the German-Spanish Astronomical Center at Calar Alto for their
friendly support during the observations, {and the referee of this
paper, Dr. G. Comte, for his helpful comments}. We are indebted to
Dr. Tom Jarrett for kindly providing photometric data from the 2MASS
survey prior to publication, which allowed us to calibrate a part of the
present sample.  We thank Dr. Uta Fritze -- v. Alvensleben, P. Anders,
J. Bicker and J. Schulz for kindly providing the GALEV models. This
research has made use of the NASA/IPAC Extragalactic Database (NED)
which is operated by the Jet Propulsion Laboratory, CALTECH, under
contract with the National Aeronautic and Space Administration.  This
publication makes use of data products from the Two Micron All Sky
Survey, which is a joint project of the University of Massachusetts
and the Infrared Processing and Analysis Center/California Institute
of Technology, funded by the National Aeronautics and Space
Administration and the National Science Foundation.
\end{acknowledgement}


\end{document}